\documentclass[12pt]{iopart}

\usepackage{color,graphicx}
\expandafter\let\csname equation*\endcsname\relax
\expandafter\let\csname endequation*\endcsname\relax
\usepackage{amsmath}
\usepackage{amsfonts}
\usepackage{amssymb}
\usepackage{ulem}

\begin{document}

\title[Methods for detecting charge fractionalization and winding numbers...]{Methods for detecting charge fractionalization and winding numbers in an interacting fermionic ladder}

\author{Leonardo Mazza$^1$, Monika Aidelsburger$^{2,3}$, Hong-Hao Tu$^2$, Nathan Goldman$^{4,5}$, Michele Burrello$^2$}

\address{$^1$NEST, Scuola Normale Superiore and Istituto Nanoscienze-CNR, Piazza dei Cavalieri 7, I-56126 Pisa, Italy}
\address{$^2$Max-Planck-Institut f\"ur Quantenoptik, Hans-Kopfermann-Str. 1,
D-85748 Garching, Germany}
\address{$^3$Fakult\"at f\"ur Physik, Ludwig-Maximilians-Universit\"at, Schellingstr. 4, D-80799 M\"unchen, Germany}
\address{$^4$Center for Nonlinear Phenomena and Complex Systems,
Universit\'e Libre de Bruxelles, CP 231, Campus Plaine, B-1050 Brussels, Belgium}
\address{$^5$Laboratoire Kastler Brossel, Coll\`ege de France,
11 place Marcelin Berthelot, F-75005, Paris, France}

\begin{abstract}

We consider a spin-1/2 fermionic ladder with spin-orbit coupling and a perpendicular magnetic field, which shares important similarities with topological superconducting wires. We fully characterize the symmetry-protected topological phase of this ladder through the identification of fractionalized edge modes and non-trivial spin winding numbers. We propose an experimental scheme to engineer such a ladder system with cold atoms in optical lattices, and we present two protocols that can be used to extract the topological signatures from density and momentum-distribution measurements. We then consider the presence of interactions and discuss the effects of a contact on-site repulsion on the topological phase. We find that such interactions could enhance the extension of the topological phase in certain parameters regimes.

\end{abstract}

\maketitle

\section{Introduction}

The experimental engineering of topological phases of matter in ultracold atomic gases~\cite{bloch13b,bloch14,esslinger14,bloch14b,mancini_2015,stuhl_2015} lays the foundations for a deeper understanding of phase transitions that transcend the Landau paradigm of symmetry breaking. In these experiments, models displaying non-local forms of order are realized in highly controllable environments, where the parameters driving the system in and out the topological phases can be tuned with wide freedom,  and where observables complementary to those of a typical solid-state experiment can be measured. These experiments have already revealed interesting properties associated with 2D topological Bloch bands: the anomalous (Hall-like) velocity, which was detected in response to an external force \cite{esslinger14,bloch14b}, the topologically invariant Chern number  \cite{bloch14b}, and chiral edge currents   \cite{bloch14,mancini_2015,stuhl_2015}.

Following these advances, an important objective would be to probe the edge modes of 1D topological systems, which typically appear at zero energy and  exhibit charge fractionalization. In particular,  identifying  an observable that unambiguously signals their presence in experiments  constitutes a remarkable challenge. Detecting the properties of zero-energy edge modes would strongly complement the Zak-phase measurement, recently demonstrated with bosonic atoms in a 1D optical superlattice \cite{bloch13b}.

Several theoretical efforts have been devoted to  the design of realistic platforms hosting topological superconducting phases with ultracold fermions~\cite{sato09, dassarma11, cirac11, diehl11, kraus12, nascimbene13, kraus13, buhler14}, including number-conserving setups~\cite{kraus13, fidkowski11, sau11, cheng11, Ortiz, Iemini_2015, Lang_2015}. Here, we envision an even simpler scenario, based on the fact that similar topological edge physics can be accessed without pairing mechanisms.  Indeed, it is a generally overlooked fact that pairing interactions are not strictly necessary to mimic topological superconductors. A fundamental example is offered by the Su-Schrieffer-Heeger model~\cite{SSH}, which presents particle-hole symmetry and belongs to a non-trivial topological class of chiral Hamiltonians, namely the class BDI of the Altland-Zirnbauer classification \cite{altlandzirnbauer,ludwig08,kitaev09,Hasan2010,Qi:2011}. 
Similar one-dimensional fermionic models without superconducting interactions display topologically-protected edge modes localized at their boundary, which are Dirac-like~\cite{Turner_2011} and feature remarkable properties, such as charge fractionalization~\cite{meng13}.

The goal of this article is twofold. First, we propose a route to mimic the physics of topological superconductors using state-of-the-art ultracold fermionic experiments. We exploit a simple idea: employing a two-leg ladder to double the fermionic species, in such a way that the legs are respectively associated with effective holes and particles~\cite{loss13,klinovaja13}, see Fig.~\ref{fig:hamiltonian}. This geometry is indeed well-suited for  experiments  on ultracold gases, as it has already been realized for atoms trapped in optical lattices, either in physical ladder geometries~\cite{bloch14}, or exploiting internal degrees of freedom as an artificial dimension~\cite{mancini_2015, stuhl_2015, Boada_2012, Celi:2014, Barbarino_2015}.
We characterize the topological properties of this model, both in the absence and presence of interactions, finding several affinities with previous studies of 1D topological interacting fermionic systems~\cite{Barbarino_2015, Gangadharaiah_2011, fisher11, sela11, hassler12, Sticlet_2014, cobanera15, franz15}.
We argue that such systems constitute a useful playground, not only to examine the appearance of symmetry-protected edge modes, but also to study the role of contact interactions, which may be tuned to drive transitions between trivial and topological phases. This analysis of the Hubbard repulsion extends to spin 1/2 fermions previous studies about the effect of interaction in generalizations of the SSH model \cite{li2013,grusdt2013}. 

As a second objective, we analyze in detail how topological signatures might be directly observed in such interacting systems. 
Beside the detection of fractionalized edge-modes, we also focus on the winding number associated with the expectation value of the spin, which provides a good detection tool for topological phases also in the presence of interactions. In this way, we extend to a one-dimensional model in the topological class BDI the techniques developed to reveal the topology of cold-atom realizations of the two-dimensional Haldane model \cite{alba11,goldman13} and other two-dimensional topological systems \cite{pachos13,burrello13,pachos14,alba15}.
For both these observables, we examine the effect of a trapping potential, which sets soft boundaries to the system (usually considered to alter the observation of edge physics). Our schemes are based on the direct observation of the atomic cloud or on time-of-flight measurements: in both cases they are extremely robust to such confinement.

This article is organized as follows.
In Sec.~\ref{sec:Model} we introduce the model and provide an intuitive description of its symmetries.
In Sec.~\ref{sec:Topological} we focus on the non-interacting model at half-filling and thoroughly characterize the topological insulator that is reached for a certain range of parameters. Furthermore, methods to detect unambiguous signatures of the topological properties are proposed, based on the density profile and the momentum-distribution of the gas.
In Sec.~\ref{sec:interactions} we study the role of interactions and characterize the related interacting topological insulator.
In Sec.~\ref{driving} we describe a possible physical realization of the model based on laser-assisted tunnelings and in Sec.~\ref{sec:conclusions} we present our conclusions. Finally \ref{app:spin} presents a detailed analysis of the  non-interacting model, its order parameter and spin winding number.

\section{The model}\label{sec:Model}

\begin{figure}[t]
 \centering
 \includegraphics[width=0.7\columnwidth]{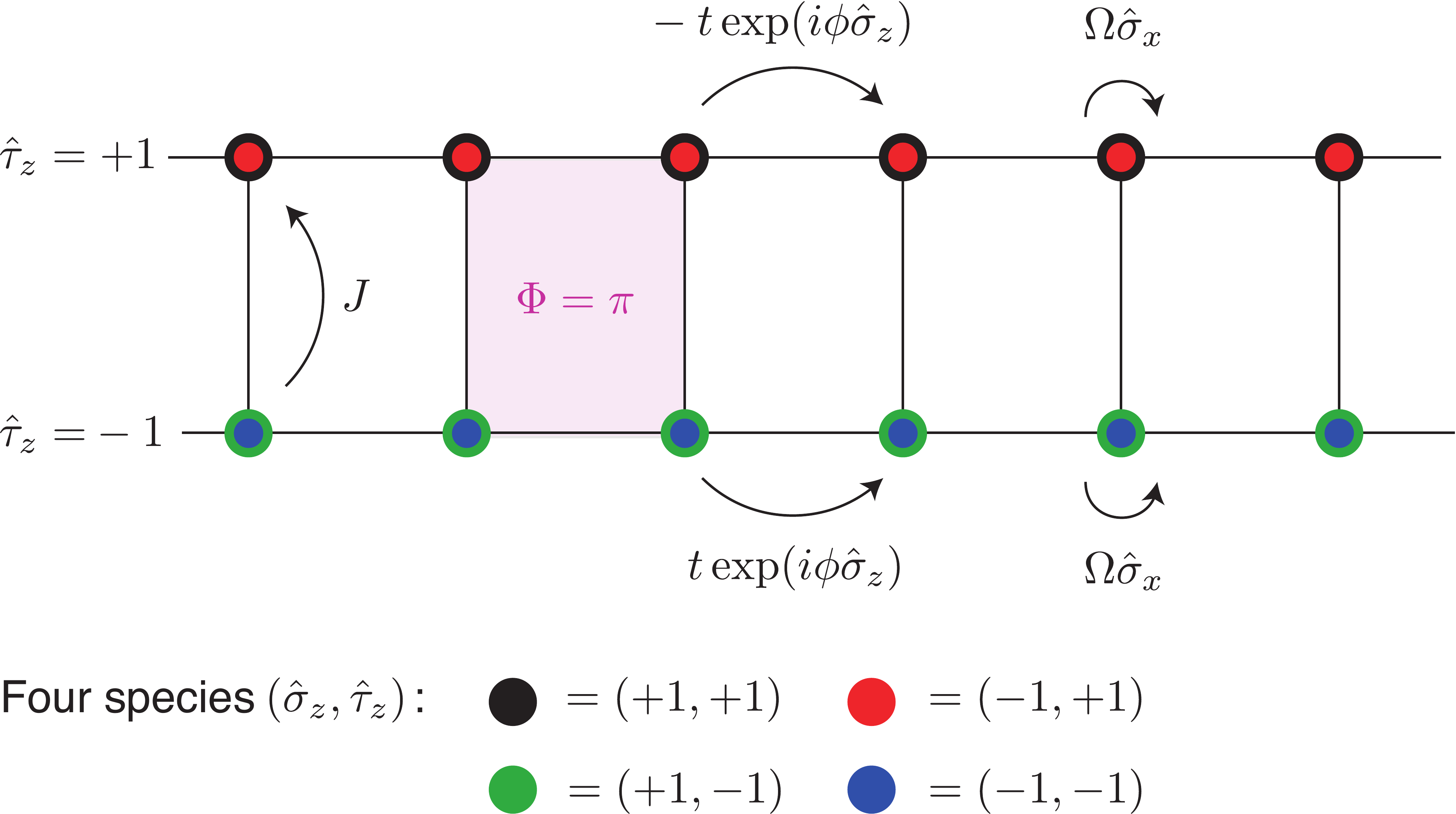}
 \caption{Schematic representation of the non-interacting model. The system is engineered in such a way that the two chains in the ladder present opposite kinetic energies. This is obtained through the introduction of a $\pi$ flux in each ladder plaquette. Here, the pseudo-spin $\tau$ refers to the two legs of the ladder, while the spin $\sigma$ is associated with the internal states (``spin") of the atoms, see Eq.~\eqref{Hamspace}. 
 } \label{fig:hamiltonian}
\end{figure}

We consider a spinful fermionic ladder in the presence of  external gauge potentials as depicted in Fig.~\ref{fig:hamiltonian}.
Here the two legs of the ladder are associated with a pseudo-spin $\tau_z$ and the lattice sites along the main axis ($x$ direction) are labelled by $r=1,...,L$.
We introduce the 4-component fermionic operator $\hat a_r$, defined on the lattice site $r$,  which acts on both the pseudo-spin $\tau_z$  and the spin
related to the two internal states of the fermion $\sigma_z$ (two commuting sets of
Pauli matrices $\tau_{\alpha}$ and $\sigma_\alpha$ are used to describe these degrees of freedom).
The Hamiltonian describing the ladder system is taken to be of the form ($\hbar =1$)
\begin{equation} \label{Hamspace}
 \hat H_0 =  \sum_r  \left\{
 t \left( \hat  a^\dag_r \tau_z e^{i\frac{B}{2} \sigma_z} \hat a_{r+1} + \text{h.c.} \right)
 +  \hat a^\dag_r \left( \Omega \sigma_x
 + J  \tau_x  + \mu \tau_z +\mu_0 \right) \hat a_r  \right\},
\end{equation}
and is schematically represented in Fig.~\ref{fig:hamiltonian}.
The first term in Eq.~\eqref{Hamspace} describes the intra-chain tunneling along the $x$ direction, with hopping amplitude $t$ and spin-dependent Peierls phase-factor $\exp (i B \sigma_z/2)$, which represents a ``spin-orbit coupling" analogous to those already realized in fermionic \cite{mancini_2015} and bosonic \cite{stuhl_2015} chains.
Note that, because of the $\tau_z$ factor, the two chains have opposite kinetic energy; consequently, the motion around each plaquette of the ladder
acquires a $\pi$-phase independently of the $\sigma_z$-spin component.
The second term describes an on-site spin-flip term with amplitude
$\Omega$, the inter-chain tunneling with amplitude $J$, the potential difference between the two chains $\mu$, and the overall chemical potential $\mu_0$.

In the following we will elaborate on the fact that
this system effectively reproduces some of the physical features of topological superconductors (TSC) without recurring to any physical pairing mechanism. This analogy is based on the idea that atoms in the first chain ($\tau_z=+1$) can be identified with conduction electrons of a generic 1D superconducting model, $\hat a^{\dagger}_{r;\,\tau_z=+1}\equiv \hat c^{\dagger}_{\text{elect}}(r)$, whereas those in the second chain ($\tau_z=-1$) can be identified with its holes, $\hat a^{\dagger}_{r;\,\tau_z=-1}\equiv \hat c^{\dagger}_{\text{hole}}(r)$. In this picture any tunneling from one chain to the other constitutes an effective pairing interaction, i.e. $\hat a^{\dagger}_{r;\,\tau_z=+1}\hat a_{r;\,\tau_z=-1} \equiv \hat c^{\dagger}_{\text{elect}} \hat c_{\text{hole}} \; { \approx} \; \hat c^{\dagger}_{\text{elect}} \hat c^{\dagger}_{\text{elect}}$, where the last equality is justified by the Bogoliubov-De Gennes treatment of the superconductor. Within this parallelism we interpret the four-band Hamiltonian \eqref{Hamspace} as a Bogoliubov-de Gennes Hamiltonian in the superconducting picture. Specifically, the particle-hole symmetry,
which plays a key role in the physics of TSCs, is here represented by a swap of the two chains, $C=\tau_y\sigma_y$, which have opposite kinetic energy in the same way as particles and holes do. Such mapping, though, must be seen only as an analogy, since the number of degrees of freedoms in the system \eqref{Hamspace} is doubled with respect to the superconducting wire and this has important physical consequences, as will be discussed in the following. Finally, note that Hamiltonian~\eqref{Hamspace} is unitarily related to those considered in the four-wire setup of Refs.~\cite{loss13,klinovaja13} and may have a relevance also for the study of electronic gases.

The model in Eq.~\eqref{Hamspace} can be realized using cold atoms trapped in an optical lattice. 
We present here an overview of the experimental proposal and refer the interested reader to Sec.~\ref{driving} for a detailed analysis of the implementation of the model.
Let's start considering a two-dimensional setup.
The realization of a spin-dependent intra-chain tunneling, described by the first term in Eq.~\eqref{Hamspace}, is particularly challenging, as it requires a subtle control over the hopping amplitudes.
This effect could be engineered by exploiting the laser-induced-tunneling methods implemented in recent experiments~\cite{bloch14b,bloch14,bloch13,ketterle13}. Specifically, we propose to achieve this task by combining a spin-dependent staggered potential with large energy offset $\Delta$ between neighboring sites, inhibiting the bare hopping along the $x$ direction, together with an onsite energy modulation set at the resonant frequency $\omega =\Delta$. The spin-dependent staggered potential is chosen to be opposite for the two internal states, i.e. $V_{\text{stag}}(r)=(-1)^{r} (\Delta/2) \sigma_z$, which can be realized by considering an appropriate anti-magic wave-length~\cite{gerbier10}; this choice is motivated by the fact that the resonant  modulation will then generate effective tunneling matrix elements of the desired form $t_{\text{eff}} (r)= t \exp [i \phi (r) \sigma_z]$, see Refs.~\cite{bloch13,kennedy2013,goldman14b}. In order to make the Peierls phase-factors  \textit{constant}  over the whole lattice, i.e. $\phi (r) = B/2$, we propose to modulate the lattice with two pairs of lasers; such a configuration allows to address individual links independently~\cite{bloch14b}, hence realizing the desired Peierls phase factors on all links, (see Section \ref{driving}).
Using additional fields resonant with the energy difference between the two spin-states a tunable onsite spin-flip term $\Omega\sigma_x$ can be realized.
Finally, the two-leg ladder can be isolated using an additional superlattice, or a light-intensity mask~\cite{corman14,corman14b}. 

\section{Topological phases in the non-interacting system}\label{sec:Topological}

\begin{figure}[t]
\centering
 \includegraphics[width=0.7\columnwidth]{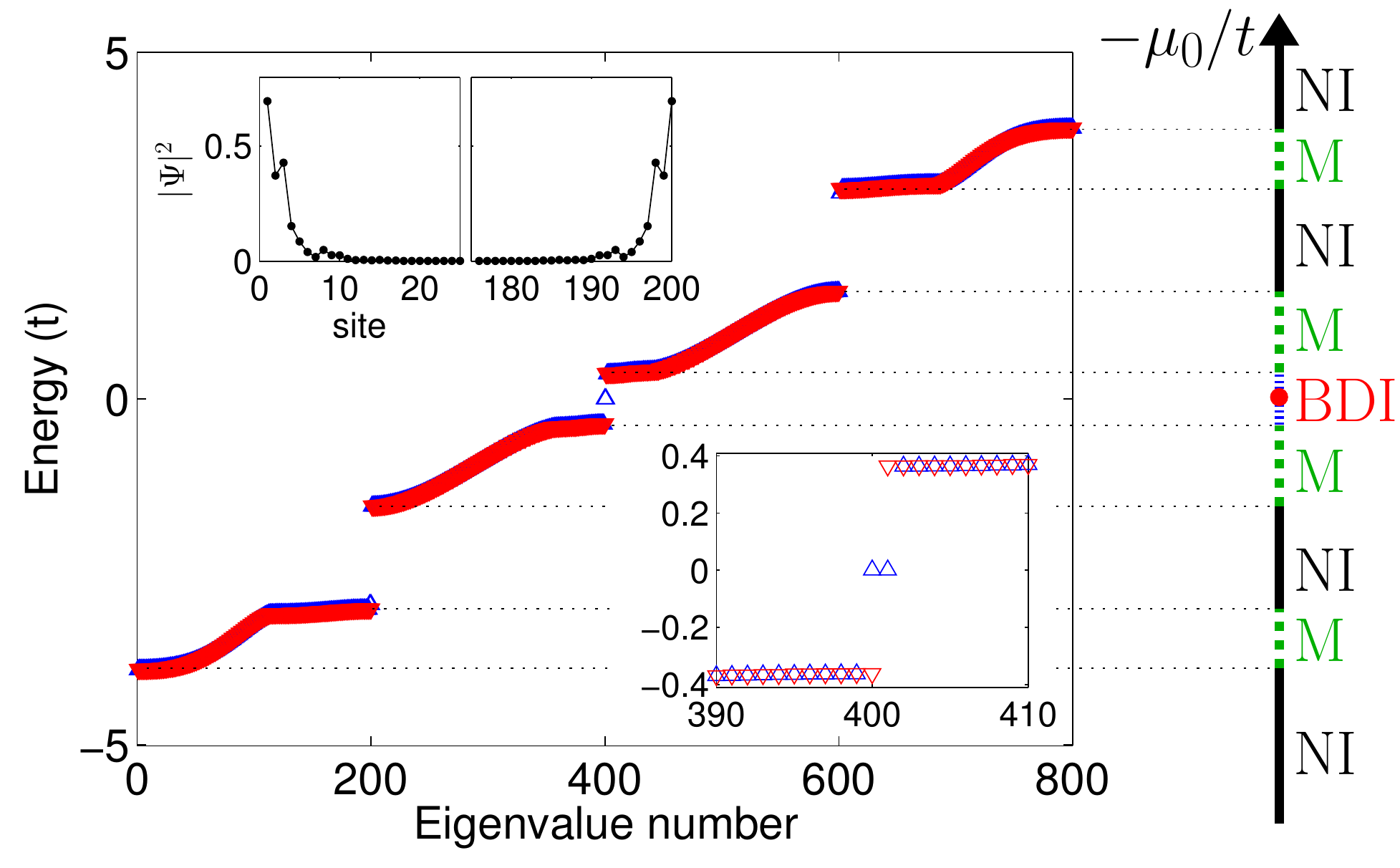}
 \caption{
 Phase diagram of Hamiltonian~\eqref{Hamspace}.
 (Left): Energy of the eigenmodes of Hamiltonian~\eqref{Hamspace} for periodic boundary conditions (red lower triangles) and open boundary conditions (blue upper triangles) for the parameters $B = \pi/2$, $J/t = 1.0$, $\Omega/t = 1.75$, $\mu/t = 0.5$, $\mu_0/t = 0$; the system size is $L=200$. The lower-right inset zooms into the zero-energy region and shows the existence of two zero-energy modes for the open system. The upper-left inset shows the squared modulus of the wavefunctions of these two modes, which are localised at the edges.
 (Right): The phase diagram as a function of $\mu_0/t$ is derived from the previous spectrum: it alternates between normal insulating phases (NI) and metallic ones (M). For half filling, thus at density $\rho = 2$ ($\mu_0=0$), the system is in the topological regime (BDI).
 }
 \label{fig:modes}
\end{figure}

Hamiltonian \eqref{Hamspace} is characterized by four energy bands;  as shown in Fig.~\ref{fig:modes}, by varying the filling of the ladder, and thus the chemical potential $\mu_0$, the system is driven through a series of metallic and insulating quantum phases (we consider in this article only the case of zero temperature). For half filling, corresponding to the case where $\mu_0=0$ and the particle density is $\rho \equiv N/L = 2$  ($N$ is the number of fermions), the single-particle Hamiltonian shows both the particle-hole symmetry we sought for, defined by the operator $C=\tau_y\sigma_y$
and an additional time-reversal symmetry, $T=\sigma_x$, which bring the system into the topological class BDI (see~\ref{app:spin} for more details). This class includes, for example, the Su-Schrieffer-Heeger (SSH) model and, according to the periodic table of topological insulators and superconductors \cite{kitaev09,ludwig08}, it may present topological phases with zero-energy modes. Specifically, our model displays a non-trivial topological insulating phase for $\Omega_{c,1}<\Omega<\Omega_{c,2}$, where $\Omega_{c,i}$ are defined, for $B<\pi$, as:
\begin{equation}
 \Omega^2_{c,1} \equiv J^2+\left( \mu - 2\cos(B/2)\right)^2 \,; \qquad
 \Omega^2_{c,2} \equiv J^2+\left( \mu + 2\cos(B/2)\right)^2\,.
\end{equation}
We find that the topological phase is surrounded by two topologically trivial phases. For $\mu_0=0$, trivial and topological phases are distinguished by a topological order parameter $\mathcal{W}$ which takes the respective values of $+1$ and $-1$ (see the Appendix and in particular Eq.~\eqref{winv} for the definition which is based on the technique developed in \cite{sau12}).

\subsection{Fractionalized edge modes}

\begin{figure}[t]
\centering
 \includegraphics[width=0.85\columnwidth]{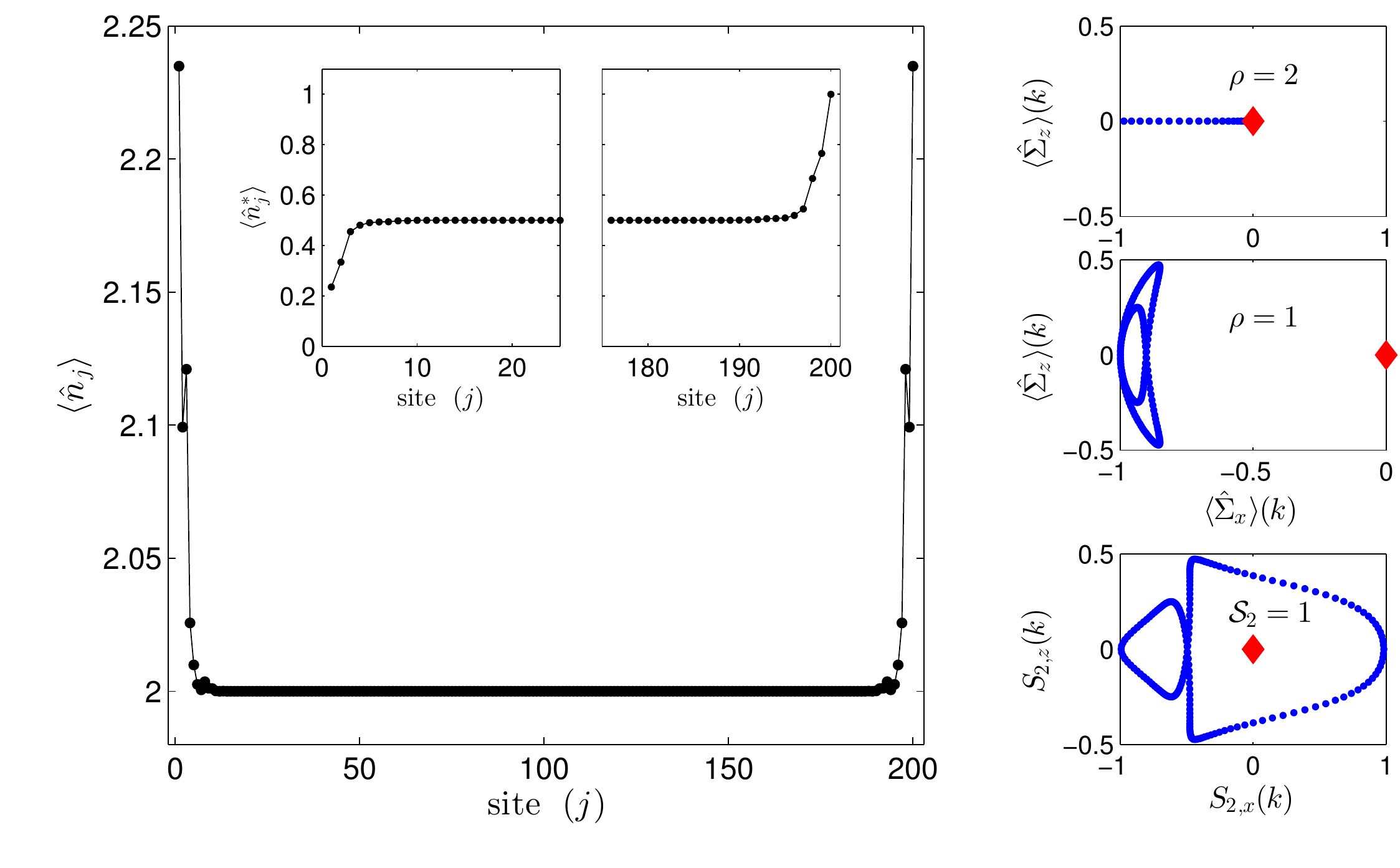}
 \caption{Charge fractionalization and spin winding number in a system with hard-walls boundary conditions.
 (Left): Density profile of the system with $N=2L+1$ particles. The simulation parameters are the same as in Fig.~\ref{fig:modes}. The inset displays the expectation value of $\hat n_j^*$.
 (Right): top and middle panels, $\langle  \boldsymbol{\hat \Sigma}(k) \rangle$ for a system with $2L$ fermions and with $L$ fermions; bottom panel, 
 $\boldsymbol{S}_2(k)/2$ for $L$ fermions artificially loaded into the second band,  see Eq.~\eqref{the_S2_equation}. The value $\mathcal W$ can be extracted only in the latter case.
 } \label{fig:profile}
\end{figure}

In the topological phase, two zero-energy fermionic modes appear in ladders with open boundary conditions, as showed in Fig.~\ref{fig:modes}. These modes are exponentially localized at the ends of the system and have important consequences on the density distribution characterizing the topological insulating phase when $N=2L+1$ fermions are introduced in the system with hard-wall boundary conditions, as displayed in Fig.~\ref{fig:profile}.
We observe that such modes are described by Dirac operators and they are not Majorana modes as it would be expected in the superconducting analog wire.
The figure shows that the density in the bulk of the system indeed corresponds to the expected value $\rho=2$.
Moreover, analogously to the SSH model, a charge $1/2$ is exponentially localized at each boundary.
This important signature of charge fractionalization can be suitably identified through the expectation value of the operator $\hat n^*_j = \sum_{m=1}^j (\hat n_m-2)$ where $\hat n_m = \hat a_m^\dagger \hat a_m$, 
which detects the excess density with respect to the bulk value $\rho = 2$.
As illustrated in the insets of Fig.~\ref{fig:profile}, an overall excess density of $1/2$ is localised within a few sites from the left  and right edges of the sample.

\begin{figure}[t]
\centering
\includegraphics[width=\columnwidth]{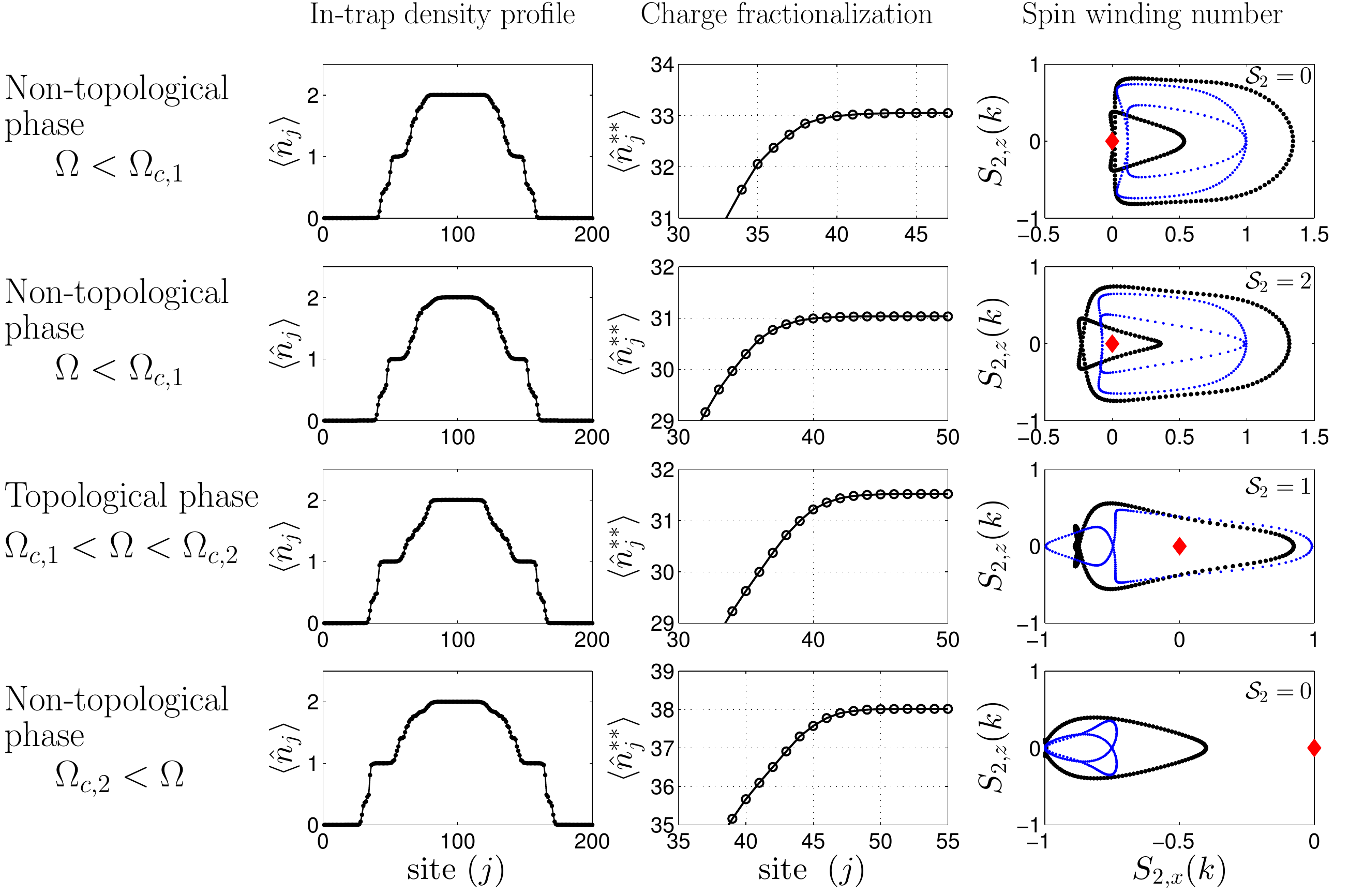}
\caption{
Charge fractionalization and winding number in a system with harmonic confinement for topological and non-topological phases.
We consider a trap with $\bar w/t = 0.03$ for $B = \pi /2$, $J/t=1$ and $\mu/t = 0.5$, for which $\Omega_{c,1}/t \simeq 1.35 $ and $\Omega_{c,2}/t \simeq 2.16$.
Different rows refer to different values of
$\Omega/t$, from up to down, $ 0.85$, $1.1$, $1.75$, $2.5$.
The number of particles is chosen to have a density $\rho=2$ in the center of the trap.
The first column displays the density of the system $\langle \hat n_j \rangle$.
The second column displays $\langle \hat n^{**}_j \rangle$.
The third column shows the winding number relative to the second band $\boldsymbol S_2 (k)$ computed for a system with trap (black) and without a trap (blue).
}
\label{fig:trap}
\end{figure}

Such a signature can be observed even in the presence of a harmonic confinement, described by the following contribution to the  Hamiltonian: $\hat H_{\rm tr} = \sum_r w_r \hat a_r^\dagger \hat a_r$, with $w_r = \bar w (r-L/2)^2$.
The effect of an external potential can be understood in a  Thomas-Fermi approach as a space-dependent chemical potential $\mu_0(r)$. Due to the four energy bands, the system has three insulating phases for intermediate fillings, and, in the presence of a harmonic trap, this yields a typical wedding-cake structure with integer density plateaus (see Fig.~\ref{fig:trap}, first column). 
Remarkably, even in the presence of the trapping it is possible to identify the fractionalized modes, as we see next. These zero-energy modes extend in the intermediate metallic region between one trivial plateau $\left(\rho=1,3 \right)$ and the topological one $\left(\rho=2 \right)$, up to exponential corrections. 

In Fig.~\ref{fig:trap}, second column, we show the expectation value of $\hat n^{**}_j = \sum_{m=0}^{j} (\hat n_{L/2+m} -1)$, 
which is particularly suited for the detection of fractionalized edge modes in cases where the density in the center of the trap is $\rho=2$. Moving from the center of the trap to the next plateau ($\rho=1$), this operator measures the excess density with respect to the $\rho=1$ value. One can obtain either an integer (no fractional modes) or an half-integer value (presence of one fractional mode).
This is an unambiguous signature of the non-trivial topological phase (third row): indeed, in this case, the quantity $\langle \hat n^{**}_j\rangle$ becomes half-integer for values of $j$ corresponding to the distance of the $\rho=1$ plateau from the center of the trap.

As a final remark, let us stress that the problem of detecting fractionalized edge modes through a density measurement was first addressed in Ref.~\cite{Ruostekoski2002, Javanainen2003}, where this detection method relies on the optical measurement of reflected light. The recent experimental advances, however, allow for the challenging method presented above, since the feasibility of a single-atom detection for ultracold fermions in optical lattices has indeed been demonstrated~\cite{kuhr2015,zwierlein15,greiner2015}. In particular, a combination of laser cooling and fluorescence detection enables an unambiguous measurement of the occupancy of single sites for both $^{40}$K \cite{kuhr2015,zwierlein15} and $^6$Li \cite{greiner2015} gases.
This is of particular importance for the detection scheme that we are proposing,
because it could suffer from the experimental inability to fix the total number of atoms which are used in the many experimental realizations necessary to reconstruct the signal $\langle \hat n_j \rangle$. 
The novel single-atom microscopes will also allow a post-selection based on the global number of particles of the system, necessary to obtain an accurate measurement.

\subsection{Spin winding number}\label{section_spin_winding}

Another interesting signature of the topological phase is offered by the expectation value of the spin operator:
\begin{equation}
 \boldsymbol{\hat \Sigma}(k)=\frac{1}{2} \hat a^\dag_k \left(\boldsymbol{\sigma} \otimes \tau_0 \right) \hat a_k
 \label{eq:spin:operator}
\end{equation}
where $\hat a_k$ is the four-component annihilation operator in momentum space. This approach is inspired by the techniques presented for two-dimensional systems in Refs.~\cite{alba11,goldman13,burrello13,pachos13,pachos14} and for ladders in Ref.~\cite{paredes14}, where it was shown that the expectation value of the spin $\langle \boldsymbol{\hat \Sigma}(k) \rangle$ provides a good observable to identify the topological invariant (winding number) of  certain topological insulators.
In the following, we generalize this procedure to our quasi-one-dimensional ladder model and show that also here the topological invariant $\mathcal{W}$, which clearly identifies the non-trivial topological regime, can be extracted from  $\langle \boldsymbol{\hat \Sigma}(k)\rangle $.  
This is thus another example of the interesting concept that a time-of-flight measurement can detect topological order.

The Hamiltonian \eqref{Hamspace} can be written in a real form thanks to its symmetries; therefore $\langle \hat \Sigma_y(k)\rangle =0$ for each eigenstate of the system and $\langle \boldsymbol{\hat \Sigma}(k)\rangle $ always lies in the $\hat{x}-\hat{z}$ plane. 
The ``spin winding number'' $\mathcal S$ is defined as the number of times the vector $\langle \boldsymbol{\hat{\Sigma}}(k)\rangle$ encircles the origin for $k$ going from $0$ to $2\pi$ (the lattice spacing is set to $1$). 
Let us denote with $\boldsymbol S_2 ( k )/2$ the expectation value of the spin operator~\eqref{eq:spin:operator} for a state of non-interacting fermions filling completely and solely the second energy band. Remarkably, the parity of $\mathcal S_2$, which is the winding number of $\boldsymbol S_2 ( k )$, coincides with the topological order parameter of the model:
\begin{equation}
\mathcal W = (-1)^{\mathcal S_2}
\label{eq:wind:top}
\end{equation}
(see~\ref{app:spin} for a demonstration and for details on the analytical calculation of this topological index for this specific model). 
Eq.~\eqref{eq:wind:top} is analogous to those derived for several other two-dimensional models~\cite{alba11,goldman13,pachos13,burrello13,pachos14}: it relates a topological invariant to a quantity, $\mathcal S_2$, to be extracted via time-of-flight imaging.
A similar behavior was discussed in Ref.~\cite{paredes14} for a two-band generalizations of the SSH model. In our case, we stress that the second band of the model is the lowest-energy band with non-trivial topological order; this is also related to the fact that the fractionalized edge modes appear in the second bulk gap.  

We now describe how to measure $\mathcal{S}_2$, for realistic systems, even in the presence of a harmonic trap. 
The main problem is that the spin winding number has to be probed for the second band of the Hamiltonian only: in a physical realization of the topological phase, both the first and the second band are filled, and the acquired signal includes information of both. The right column of Fig.~\ref{fig:profile} shows the expectation value of the spin $\langle  \boldsymbol{\hat \Sigma}(k) \rangle$ obtained when the system with hard-wall boundaries is filled with $2L$ fermions (top panel) and with only $L$ fermions (middle panel). Both signals are not particularly interesting. If we consider the artificial situation where atoms populate the second band only (bottom panel), the spin expectation value is characterized by a winding number that reproduces the behavior of $\mathcal{W}$ and encircles the origin in the topological phase.
In the ideal case of a hard-wall confining potential, the required value of the second band can be extracted by repeating the experiment twice, at densities $\rho=1$ and $\rho=2$: the difference of the measured distributions returns the sought information
$\langle  \boldsymbol{\hat \Sigma}(k) \rangle_{\rho=2}- \langle \boldsymbol{\hat \Sigma}(k) \rangle_{\rho=1} = \frac 12  \boldsymbol { S}_2(k) $.

In the presence of a harmonic trap, the wedding cake density profile suggests that the many-body wavefunction can be roughly thought as a state where each energy band $\alpha$ is uniformly populated by $N_\alpha$ atoms ($N_1 \geq N_2 \geq N_3 \geq N_4 \geq 0$). In this case the measurement of the observable $\langle \boldsymbol{\hat \Sigma}(k)\rangle$ returns:
\begin{equation}
 \langle \boldsymbol{\hat \Sigma}(k)\rangle=\frac{1}{2}\sum_\alpha \frac{N_\alpha}{\mathcal{L}}\boldsymbol{S}_\alpha(k)
\end{equation}
where $\boldsymbol{S}_\alpha(k)$ is the expectation value of the spin calculated in the thermodynamic limit for the single particle eigenstate of the
$\alpha^{\rm th}$ energy band (see \ref{app:spin}). $\mathcal{L}$ is the discretization adopted for the Brillouin zone in the time-of-flight imaging (see, for example, \cite{alba11,burrello13}).

If we consider the case in which the density profile shows only two plateaus, the value of $\boldsymbol{S}_2(k)$ can be estimated by comparing the observed $\langle \boldsymbol{\hat \Sigma}(k)\rangle_{\rho=2}$ with that of a realization with a single plateau only, $\langle \boldsymbol{\hat \Sigma}(k)\rangle_{\rho=1}$:
\begin{equation}
 \boldsymbol{S}_2(k)=\frac{2\mathcal{L}}{N_2}\left(\langle \boldsymbol{\hat \Sigma}(k)\rangle_{\rho=2}-\frac{N_1}{N_1'}\langle \boldsymbol{\hat \Sigma}(k)\rangle_{\rho=1} \right) \label{the_S2_equation}
\end{equation}
where $N_1$ and $N_2$ are the occupations of the two bands for the state with two plateaus, and $N_1'$ is the total number of atoms in the reference state with a single plateau. All the quantities $N_1,N_2$ and $N_1'$ can be experimentally accessed and we report in the right column of Fig.~\ref{fig:trap} the comparison of the data obtained for hard wall and harmonic potentials.
Our numerical simulations confirm that even in the presence of the trap $\mathcal S_2$ is equal to $\pm 1$ in the topological phase, whereas in the trivial phases, it is either 0 or $\pm 2$ (see Fig.~\ref{fig:trap}).

Let us conclude with some information on how to measure 
$\langle \boldsymbol{\hat{\Sigma}}(k)\rangle$ through spin-resolved time-of-flight imaging \cite{alba11,goldman13,burrello13,pachos13,pachos14} in our setup. Special care is required in time-modulated systems with spin-dependent features \cite{dalibard14}, as considered in the specific proposal detailed in Sec.~\ref{driving} because spin-dependent observables can potentially undergo large and complicated micro-motion (rapid motion with a time-scale of the order of the driving period $2 \pi / \omega$), which typically alters the accuracy of measurements. In such schemes, stroboscopic measurements performed at specific times, $2 \pi / \omega \times n$ where $n$ is integer, are generally required to extract relevant information relative to the spin-dependent quantities~\cite{dalibard14}.  For the scheme detailed in  Sec.~\ref{driving}, the micro-motion can be estimated from the unitary operators $K (t)$ and $R(t)$ defined in that Section, through the method of Ref.~\cite{goldman14b}. We find that $\langle \hat \Sigma_z(k)\rangle$ is unaffected by the micro-motion; in contrast, an accurate analysis of $\langle \hat \Sigma_x\rangle $ does require a stroboscopic measurement. 
Moreover, we note that measuring the expectation value of $\hat \Sigma_x$ also necessitates a $\pi/2$ pulse, which has to be short compared to the driving period in order to probe the system stroboscopically.

\section{Interacting system}\label{sec:interactions}

\begin{figure}
\begin{center}
 \includegraphics[width=0.49\columnwidth]{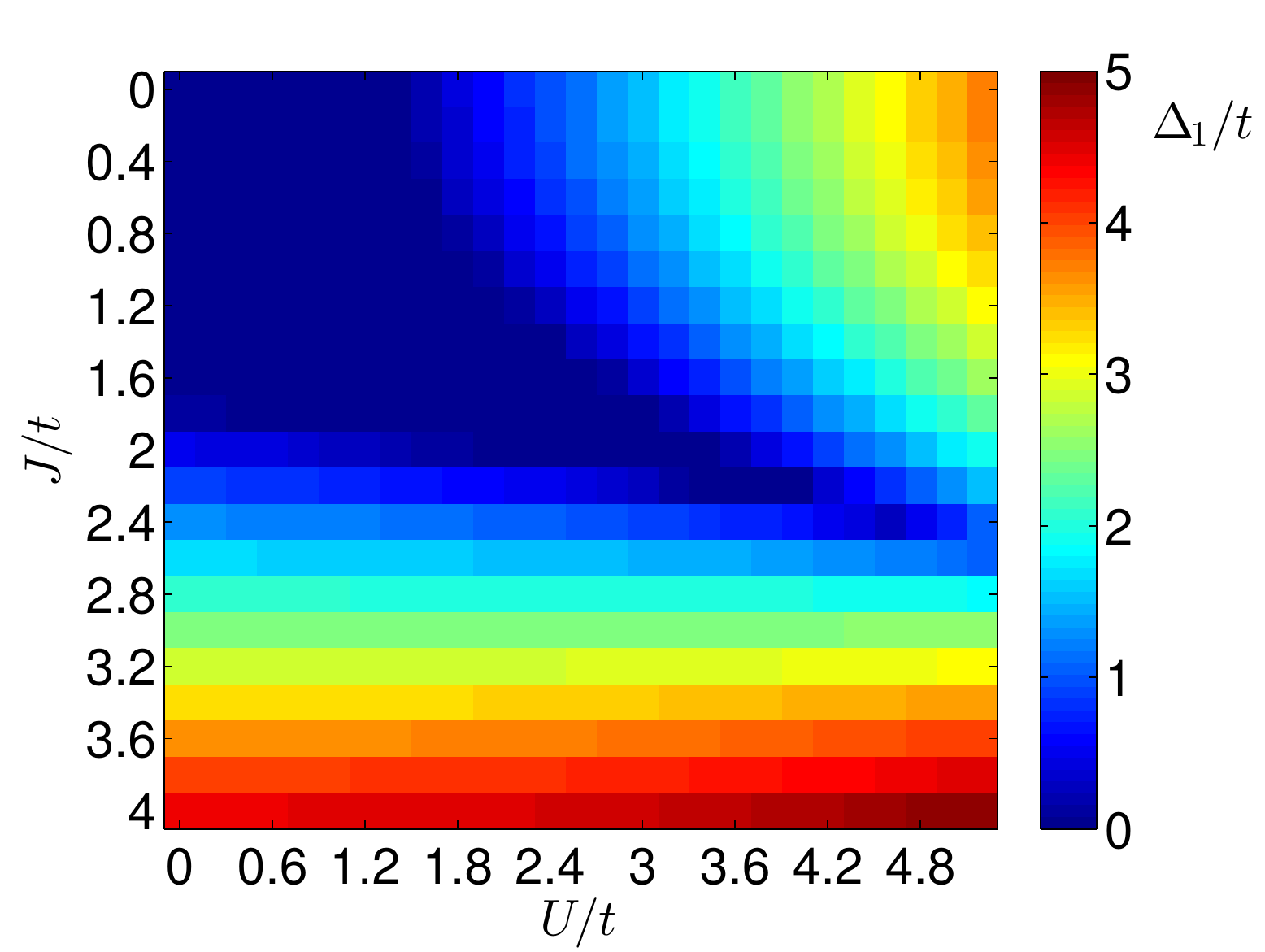}
 \includegraphics[width=0.49\columnwidth]{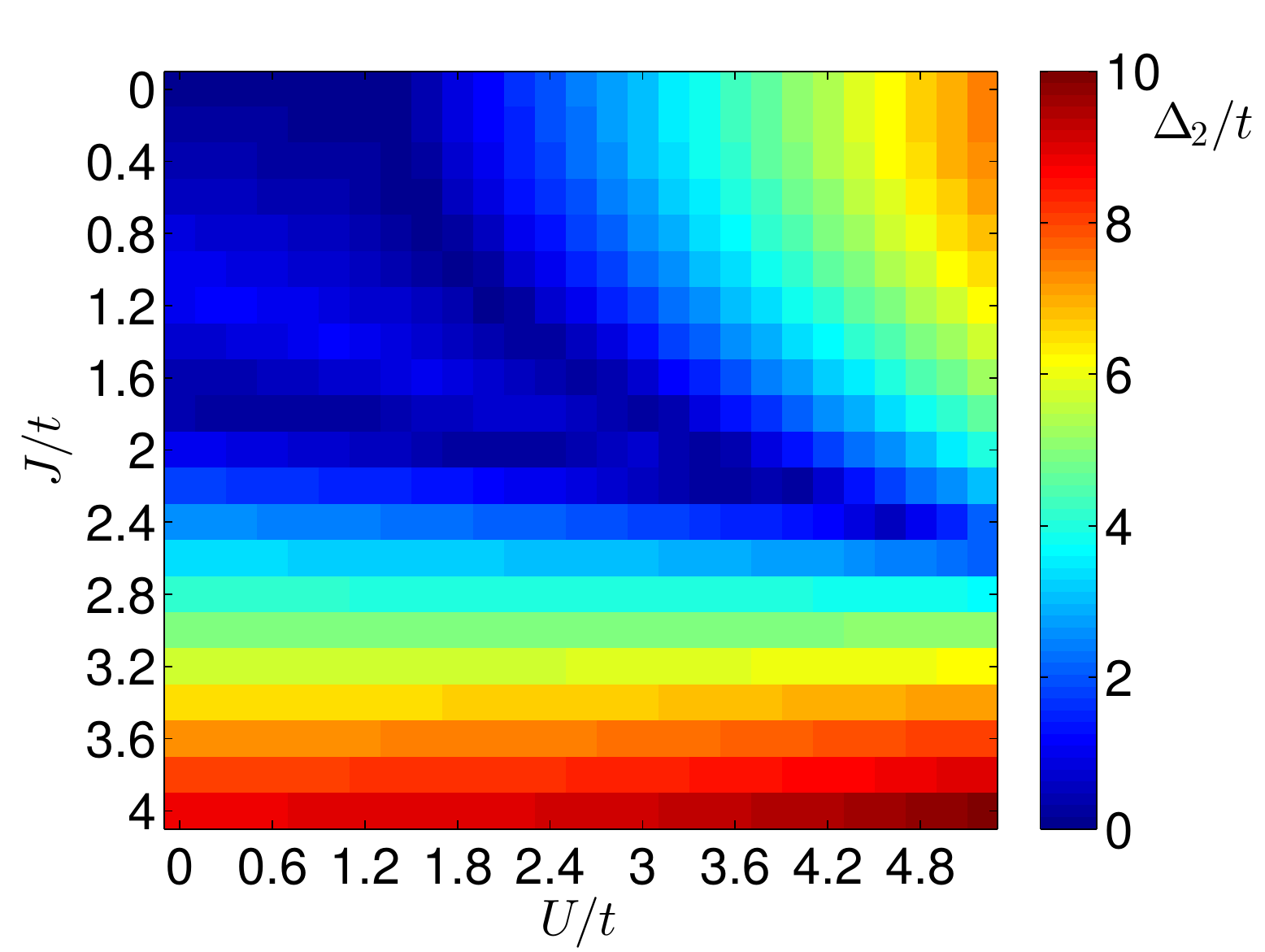}
 \end{center}
 \caption{Phase diagram of the interacting model at half-filling in the $U/t$ and $J/t$ plane. (Left) Single-particle gap $\Delta_1 / t$. (right) Two-particle gap $\Delta_2 / t$. The topological region is characterized by $\Delta_1=0$ (dark blue region in the left panel) and it is delimited by gapless regions defined by $\Delta_2=0$ (dark blue regions in the right panel) as represented schematically in Fig. \ref{fig:scheme}.
 The calculations are performed for $\Omega/t=1.8$, $B= \frac{\pi}{2} \frac{L}{L+1}$ and $ \mu/t=1$ at $L=72$ with bond dimension $D=200$.}
 \label{fig:PhaseDiagram}
\end{figure}

Let us now consider the role of interactions, with a special emphasis on the robustness of symmetry-protected topological order.
It is experimentally relevant to consider an on-site Hubbard interaction in each leg:
\begin{equation}
 \hat H_{\rm int} = U \sum_{r, \tau_z} \hat n_{r, \tau_z, \sigma_z = \uparrow}
 \hat n_{r, \tau_z, \sigma_z = \downarrow}.
\end{equation}
and to analyze the phase diagram of $\hat H_0 + \hat H_{\rm int}$ at half filling, $\rho=2$, which is characterized by the competition between the topological insulator (TI) and Mott insulator (MI) occurring in the presence of a strong contact repulsion.
We employ a density-matrix renormalization group algorithm based on a Matrix-Product State (MPS) ansatz~\cite{White_1992, Schollwoeck_2011}. We will consider systems with open-boundary conditions with $L=72$ and maximal bond dimension $D=200$.

The transition between TI and MI can be located via the analysis of the charge gap  at $N=2L$.  In particular, the single-particle gap is defined as:
\begin{equation}
  \Delta_1(N) =  E(N+1) + E(N-1)-2 E(N),
\end{equation}
where $E(N)$ is the ground-state energy of the system with $N$ fermions.
Clearly, $\Delta_1(N)>0$ for the MI because the system has a thermodynamic gap. On the other hand, the TI has zero-energy modes which ensure that $E(N-1) = E(N) = E(N+1)$ and thus $\Delta_1(N) =0$.
Unfortunately, the mere calculation of $\Delta_1$ does not permit to discriminate the TI from a generic gapless phase,  for which $\Delta_1(N)=0$ too.  We thus consider also the two-particle gap:
\begin{equation}
  \Delta_2(N) = E(N+2) + E(N-2) - 2 E(N) .
\end{equation}
 Whereas for a gapless phase $\Delta_2(N)$ is also equal to zero, for a TI it is larger than zero, signaling the gap which is protecting the phase.

Based on this discussion, we now consider a systematic study of the Hamiltonian, focusing on the competition of the two terms which are responsible for a gap opening, namely, the interaction term proportional to $U$ responsible for the MI,
and the interchain tunneling proportional to $J$.
Roughly speaking, we identify the pairing term as the one inducing the TI, since at $U=0$ the system is in a topological phase for $0<J^2 < J_{c,1}^2 \equiv \Omega^2 - (2\cos(B/2)-\mu)^2$ (see \ref{app:spin} for more detail) and the two chains decouple at $J=0$. 

\begin{figure}
\begin{center}
\includegraphics[height=0.33\columnwidth]{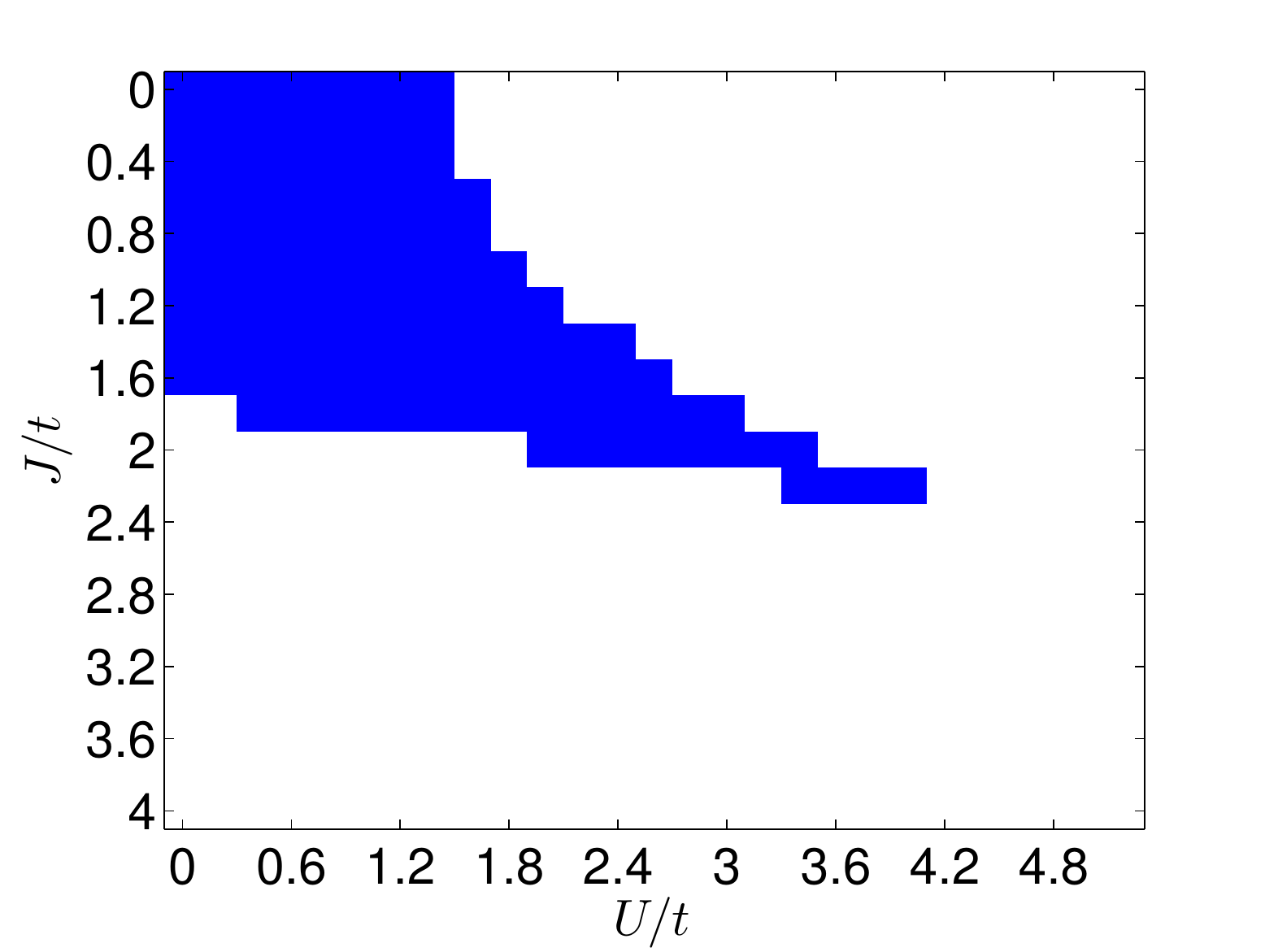}
 \includegraphics[height=0.315\columnwidth]{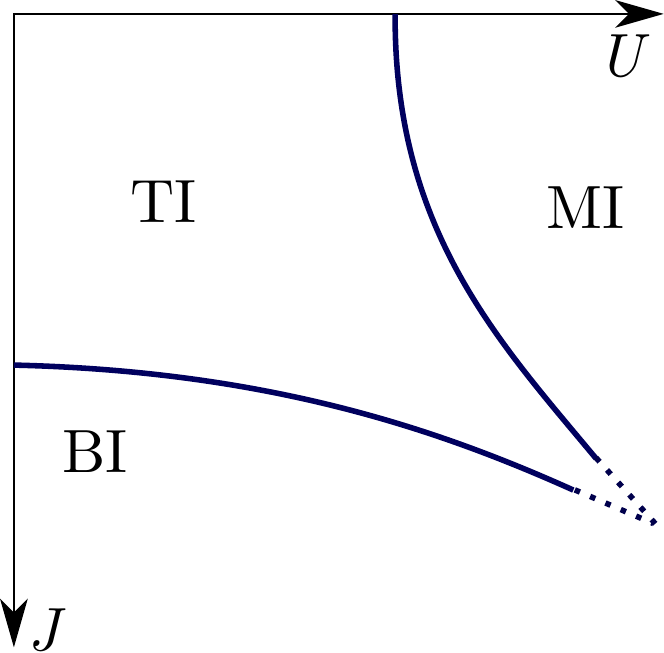}
 \end{center}
 \caption{Schematic structure of the phase diagram at half filling as a function of $J$ and $U$. (Left) The blue region represents the topological region as extracted from the numerical results presented in Fig.~\ref{fig:PhaseDiagram} (note that strictly speaking our numerics could not access the thermodynamic limit).
 (Right) Qualitative extrapolation of the phase diagram.
 Three gapped phases can be detected: topological insulator (TI), trivial band insulator (BI) and Mott insulator (MI). Blue lines represent the phase transitions. Our numerical results suggest that the MI and BI phases are adiabatically connected. The exact nature of the phase diagram in the dotted region cannot be established due to numerical limitations.} \label{fig:scheme}
\end{figure}

Fig.~\ref{fig:PhaseDiagram} presents 
the numerical results for $\Delta_1(2L)$ and $\Delta_2(2L)$ in the parameter space spanned by $U/t$ and $J/t$.
The other parameters are chosen such that at $U=0$ there is a TI, and are listed here for completeness: $\Omega/t=1.8$, $B = \frac{\pi}{2} \frac{L}{L+1}$ and $\mu/t =1$.
Calculations are limited to the size $L=72$ and a systematic scaling to the thermodynamic limit, as well as the exact evaluation of the properties of the critical lines, is beyond our numerical possibilities; additionally, the two-dimensional space is studied with a grid of $0.2$ along both axis.
Despite these limitations, the qualitative nature of the phase diagram emerges quite clearly.
Indeed, through the study of $\Delta_1$ and $\Delta_2$ we are able to identify the TI, the MI and the critical regions which separate them, resulting in the schematic phase diagram presented in Fig. \ref{fig:scheme}.
The topological region is identified with the large region where $\Delta_1 = 0$ but $\Delta_2 > 0$ whereas for the MI both $\Delta_1 $ and $\Delta_2 $ are larger than zero.
Critical regions with $\Delta_1 = \Delta_2 = 0$ separate the two insulators.
The schematic phase diagram discriminates the MI, whose appearance is driven by the on-site repulsion, from the trivial band insulator (BI), which appears instead also at $U=0$. Our investigation did not identify a phase transition between these two trivial insulating phases, which are adiabatically connected.

\begin{figure}[t]
\includegraphics[width=\columnwidth]{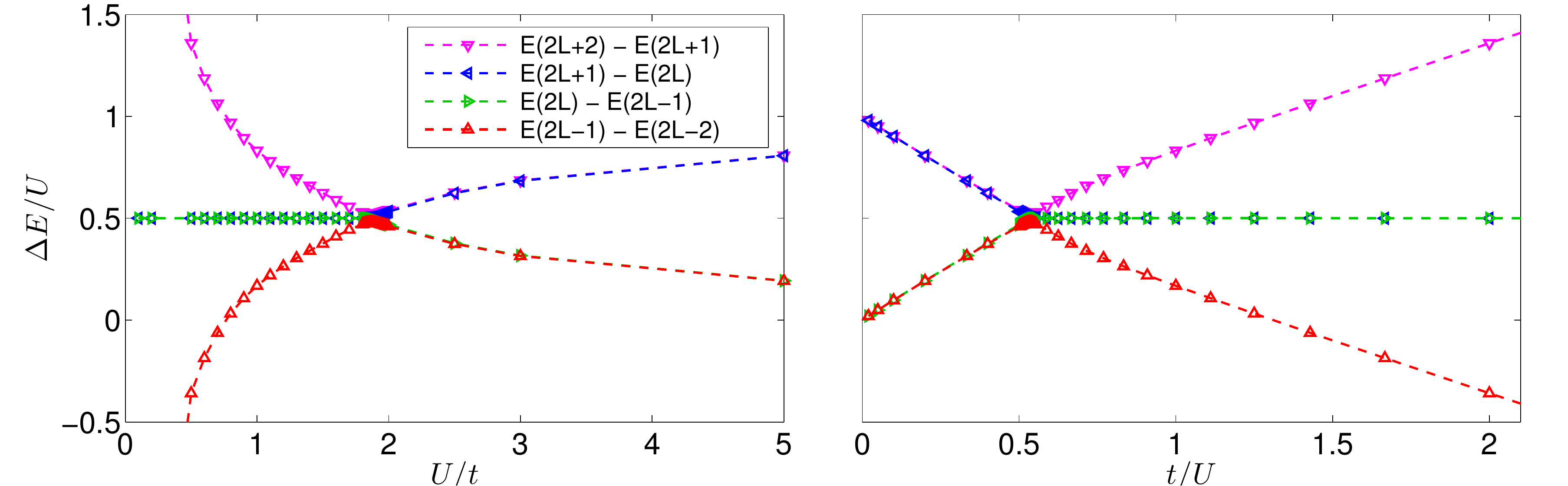}
\caption{Energy differences of the ground states of the system for several fillings around $N=2L$.
The same data are plotted as a function of $U/t$ (left) and $t/U$ (right).
The Hamiltonian parameters are $J/t = 1$, $\Omega/t = 1.8 $, $B = \frac{\pi}{2} \frac{L } {L+1}$, so that the cut corresponds to the line $J/t=1$ in the phase diagram of Fig.~\ref{fig:PhaseDiagram}.
$L = 72$ and the maximal MPS bond dimension is $D = 150$}
\label{fig:Gaps}
\end{figure}

To better analyze the transition between TI and MI,
in Fig.~\ref{fig:Gaps} we focus on the line at $J/t=1$, which entails a phase transition for $U/t=u_{cr} \sim1.9$.
We show the behaviour of the chemical potentials $E (2L+\alpha)-E(2L+\alpha-1)$ for $\alpha = +2, +1, 0, -1$ as a function of $U$.
Two qualitatively different behaviours are separated by $u_{cr}$.
For $U/t> u_{cr}$ the energy cost for adding one particle to the states with $N=2L$ or $N=2L+1$ is approximately $U$ (especially for large values of $U/t$).
Conversely, subtracting one particle from the states with $N=2L$ or $N=2L-1$ does not yield any energy gain. Thus, $\Delta_1(N=2L)$ and $\Delta_2(N=2L)$ are larger than zero and are approximately equal to $U$ and $2U$, respectively:
these are typical signatures of a MI.

For $U/t < u_{cr}$ the energy cost and gain for adding and removing  one particle to/from the state with $N=2L$ are both equal to $U/2$. We interpret this as a  signature of the fractionalization of the zero-energy modes of the TI: the charge excess $n \sim 1/2$ on top of the density plateau $\rho = 2$ (one particle per site) does cost a repulsive energy $U n$.
Since the zero-energy modes have fermionic nature, they cannot accommodate more than one particle: the energy cost for adding one additional particle to the state with $N=2L+1$ becomes significantly larger than $U/2$ (vice versa for removing one particle from the state with $N=2L-1$).
Thus, $\Delta_1(N=2L)=0$ but $\Delta_2(N=2L)>0$, signaling a TI for $U/t< u_{cr}$.

The MI extends for $U/t \gg 1$, where a perturbative expansion shows that the system can be described by a spin model in a paramagnetic phase: under the assumptions $U\gg J,t$ and $\rho=2$, one atom is trapped in each site of the two legs. 
We  thus introduce the Pauli operators $\hat \eta^i_{r,\tau}$ ($i=x,y,z$) acting on the
 local effective Hilbert space 
spanned by the two spin states $\sigma = \pm1$ of the atom located
at the site $r$ of the chain $\tau=\pm1$.
We obtain the following second-order perturbative spin Hamiltonian:
\begin{multline}
 \hat H_{\rm pert}=\sum_{r,\tau} \left[ \Omega \hat \eta_{r,\tau}^x+ J_{\rm eff}\hat \eta_{r,\tau}^z\hat \eta_{r+1,\tau}^z +J_{\rm eff}\cos B\left(\hat \eta_{r,\tau}^x \hat \eta_{r+1,\tau}^x+\hat \eta_{r,\tau}^y \hat \eta_{r+1,\tau}^y\right)\right.    \\
 \left.+J_{\rm eff}  \sin B \left(\hat \eta_{r,\tau}^x\hat \eta_{r+1,\tau}^y-\hat \eta_{r,\tau}^y\hat \eta_{r+1,\tau}^x\right)\right]  + \sum_{r,j}K_{\rm eff} \hat \eta_{r,\tau=1}^j\hat \eta_{r,\tau=-1}^j 
\end{multline}
where $J_{\rm eff}\propto t^2/U$ and $K_{\rm eff}\propto J^2/U$. 
In this regime, the term proportional to $\Omega$ dominates and the ground state of $H_{\rm pert}$ is close to a trivial product state in which all the spins are oriented in the $\hat x$ direction. We expect that such state, characterizing the MI phase, might be adiabatically connected to the trivial band insulator at $U=0$ and $J>J_{c,1}$. Our numerics does not suggest the existence of a further phase transition between the Mott and the trivial band insulating phases.  

The phase diagram in Fig.~\ref{fig:PhaseDiagram} shows that the topological region appears clearly as a thermodynamic region, within a well defined parameter regime. We emphasize that, for $1.8 \lesssim  J\lesssim2.2$, the system is in a topologically trivial phase for $U=0$, and enters the symmetry-protected topological phase when the interaction parameter $U$ is increased. Therefore the interaction is not necessarily obnoxious to the purpose of experimentally obtaining the topological phase but, on the contrary, it can also drive the system into it by shifting the position of the critical point. This has been verified also in the corresponding topological superconductor systems \cite{fisher11,sela11,hassler12}, where the addition of repulsive interactions is proven to expand the topological phase for certain ranges of the physical parameters (see also \cite{cobanera15,franz15} for related models in terms of Majorana modes). This means that, for some particular value of $J>J_{c,1}$ the presence of a repulsive interaction allows the formation of edge modes otherwise absent. A similar behavior is also observed in 2D systems with time-reversal invariance~\cite{sangiovanni13}.
Let us stress, however, that this has nothing to do with the physics of fractional Chern insulators, where interactions drive the system into distinct (strongly-correlated) topological phases. As the phase diagram clearly shows, there is only one TI phase, which is strictly equivalent to that of the non-interacting system.  Importantly, the phase diagram in Fig.~\ref{fig:PhaseDiagram} shows that interactions have a non-trivial role in tuning the system in and out the TI phase.

\begin{figure}[t]
\centering
  \includegraphics[width = 0.49\columnwidth]{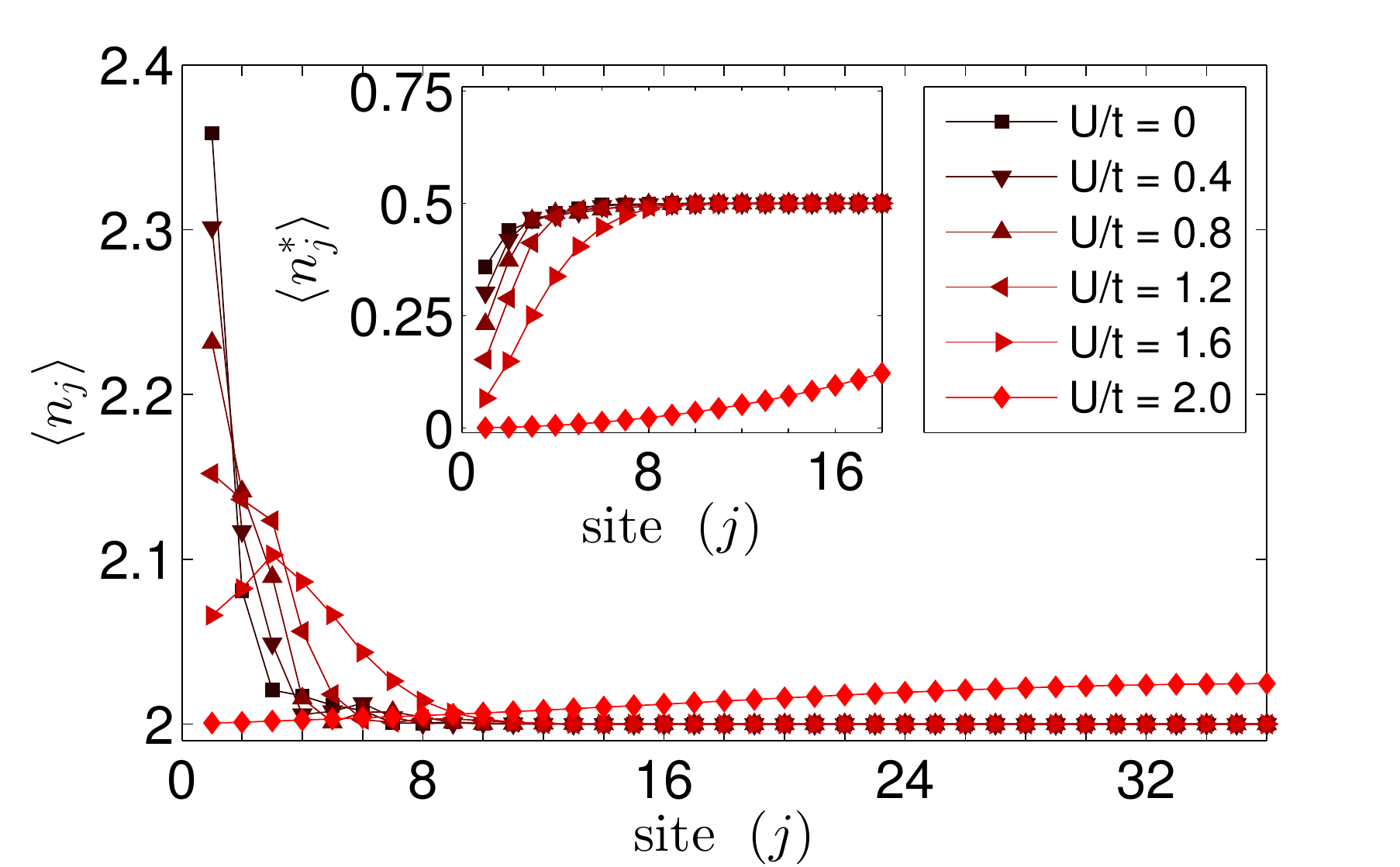}
  \includegraphics[width = 0.49\columnwidth]{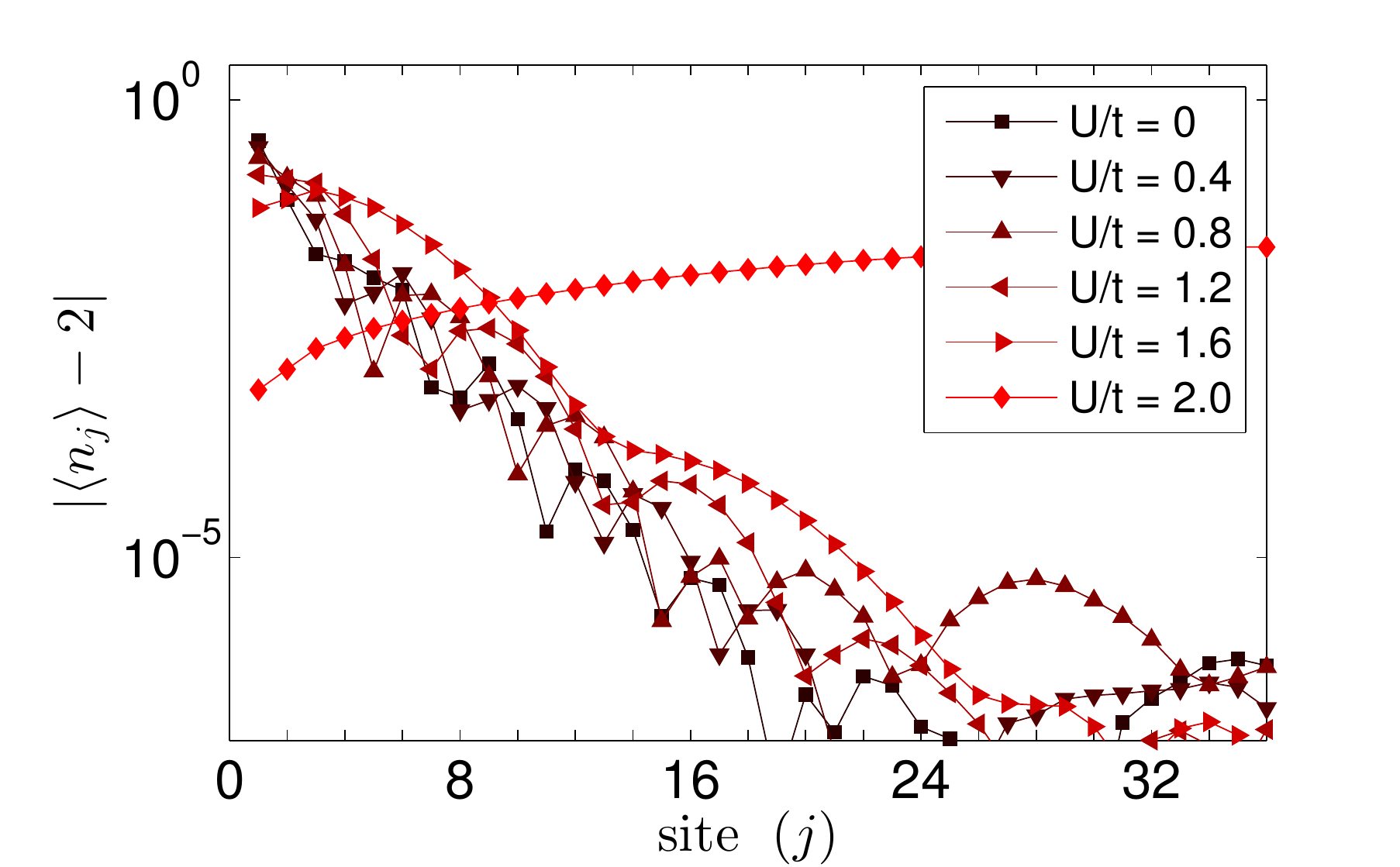}

  \includegraphics[width = 0.49\columnwidth]{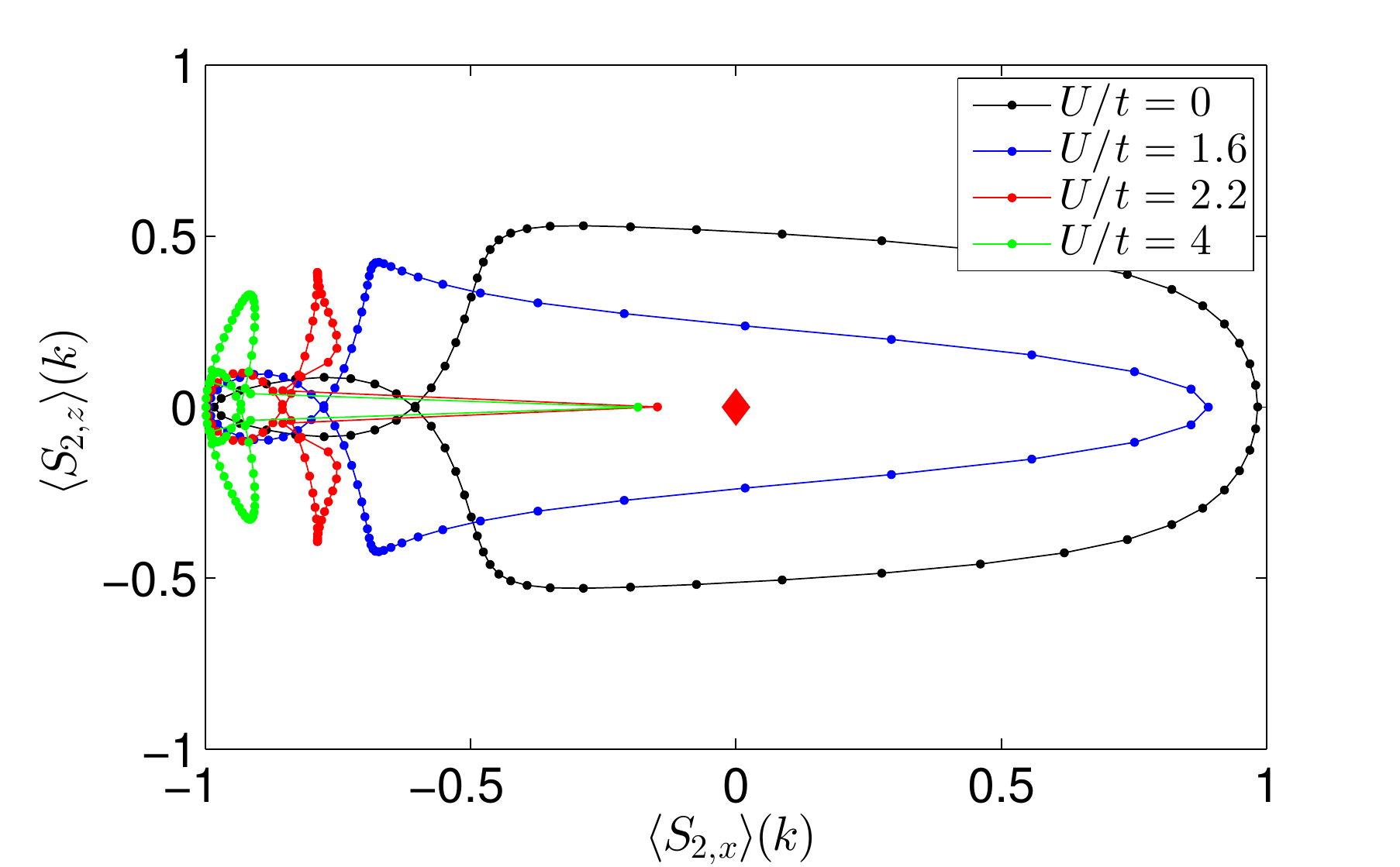}
  \includegraphics[width = 0.49\columnwidth]{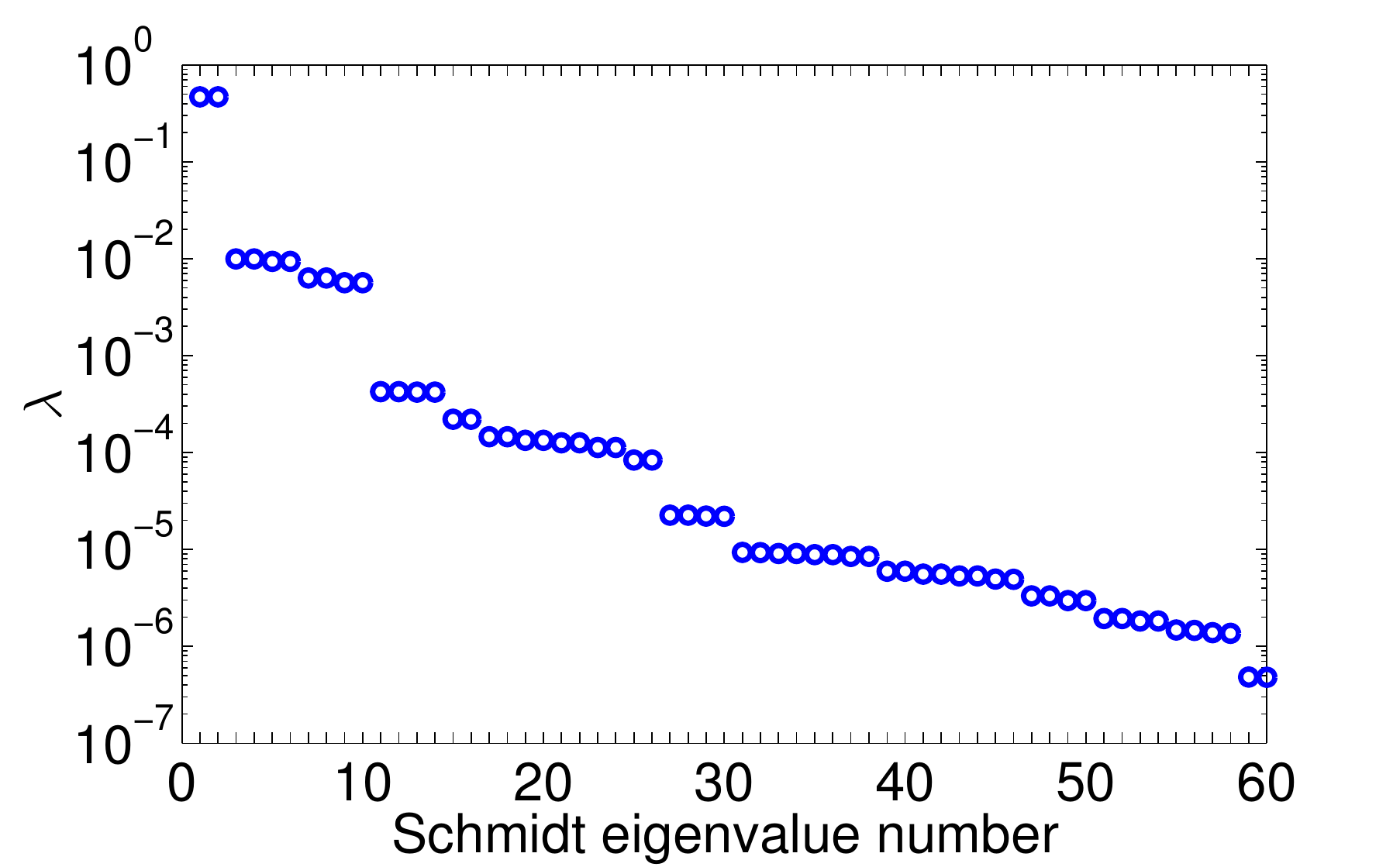}

\caption{
Properties of the interacting topological phase for $J/t = 1$, $\Omega/t = 1.8 $, $B = \frac{\pi}2 \frac{L}{L+1}$: all the analyzed systems are on the line $J/t=1$ in the phase diagram in Fig.~\ref{fig:PhaseDiagram}. $L = 72$ and MPS bond dimension $D = 200$.
(top, left) Density profiles $\langle \hat n_j \rangle$ and $\langle \hat n^*_j\rangle$ of the gas with $U/t =1$ for $N=2L$ and $N=2L\pm 1$. The data clearly show the presence of localised and fractionalised edge modes.
(top, right) The plot of $|\langle \hat n_j \rangle-2|$ highlights the localisation of the edge modes.
(bottom, left) Spin winding associated to the second band of the model $\boldsymbol S_2 $ for several values of the interaction, within and without the TI.
(bottom, right) Entanglement spectrum ($60$ largest eigenvalues) for $N=2L$ and $U/t=1$.}
\label{fig:schmidtspectrum}
\end{figure}

In order to further clarify this last point, we now investigate in more detail the properties of the topological phase in the interacting system.
Numerical investigations reported in Fig.~\ref{fig:schmidtspectrum} show that the signatures of the non-interacting TI persist in the presence of interactions.
First, the density profile of the gas allows for a clear identification of the presence of fractionalized edge modes located at the boundaries of the ladder via the computation of $\langle \hat n_j \rangle$ and $\langle \hat n^*_j \rangle$. Indeed, for $U/t \leq u_{cr}$, Fig.~\ref{fig:schmidtspectrum} shows that $\langle \hat n^*_j \rangle$ saturates to $0.5$ within few sites, which is strongly different from the behaviour for $U/t \geq u_{cr}$.
It is interesting to observe that within the topological region the localisation length of the edge modes has a weak dependence on $U/t$.
Second, the system displays also within the interacting region a non-zero winding number associated with the second band of the system. As in an interacting system bands are not well defined, the winding number is computed by subtraction of the data relative to $\rho = 1$ to those relative to $\rho =2$ (see similar discussion in Sec.~\ref{sec:Topological}). This robustness of the spin winding number against local interaction is consistent with similar results in two-dimensional systems \cite{alba15}.
Finally, on a more abstract side, the analysis of the Schmidt spectrum presents the robust two-fold degeneracy of symmetry-protected topological phases \cite{pollmann10}.

\section{Physical realization of the model} \label{driving}

\begin{figure}[t]
 \includegraphics{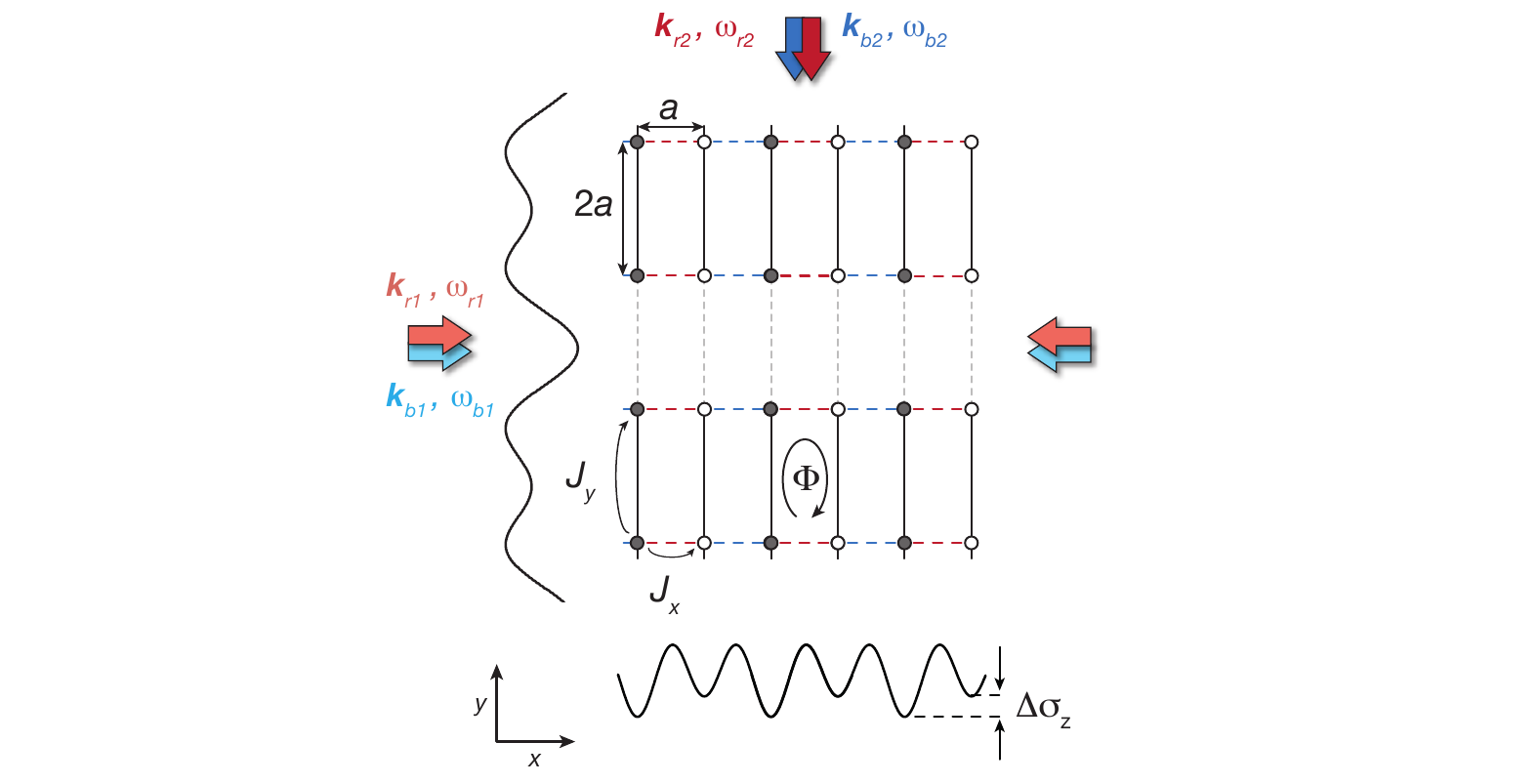}
 \caption{ Schematic drawing of the proposed experimental setup. The 2D lattice configuration consists of a spin-independent superlattice along $y$ to isolate individual ladders from each other and a superlattice potential along $x$, which creates a spin-dependent energy offset $\Delta \sigma_z$ between neighboring sites in order to inhibit tunneling. Tunneling is then restored resonantly with two pairs of beams denoted as $r$ and $b$ following the scheme introduced in Ref.~\cite{bloch14b}. Each of the pairs consists of a standing wave along $x$ and a running-wave along $y$. For $\omega_{r/b}=\omega_{1,r/b}-\omega_{2,r/b}=\pm \Delta$ and $\mathbf{q_{r/b}}=\mathbf{k}_{1,{r/b}}-\mathbf{k}_{2,{r/b}}=(1, 1) \cdot \pi/(2a)$ an effective flux $\Phi=\pi$ is realized with spin-dependent complex tunneling-matrix elements, here $a$ is the lattice constant of the potential along $x$.}\label{fig:realization}
\end{figure}

The physical realization of the ladder system in Eq.~\eqref{Hamspace} can be obtained by extending the 2D setup elaborated and realized in Ref.~\cite{bloch14b}. The present proposal builds on a 2D optical superlattice subjected to a well-designed time-modulation as displayed in Fig.~\ref{fig:realization}. 
 Along the $y$ direction a superlattice potential is used to partition the lattice into a 1D array of isolated ladders. Hopping between the two legs of the ladder corresponds to transitions $\hat \tau_z=-1 \leftrightarrow +1$, see Fig.~\ref{fig:hamiltonian}. The main challenge in realizing the Hamiltonian in Eq.~\eqref{Hamspace} consists in engineering the spin-dependent complex matrix elements for tunneling processes taking place along the legs of the ladder. In the following, we will show that this can be achieved by combining a spin-dependent superlattice potential $x$ (Fig.~\ref{fig:realization}), which introduces a spin-dependent energy offset $\Delta$ between neighboring sites, together with the space-dependent time-modulation of the lattice discussed in Ref.~\cite{bloch14b}.

We start by considering the time-independent part of the system, which can be described by the 2D tight-binding Hamiltonian 
\begin{align} \label{ham_lab}
 \hat H_0 =& -J_x \sum_{m,n} \left( \hat a_{m+1,n}^\dag \hat a_{m,n} + \text{H.c.}\right)  + \\
 &- J_y \sum_{m,n} \left( \hat a_{m,n+1}^\dag \hat a_{m,n} + \text{H.c.}\right)
 + \frac{\Delta}{2} \sum_{m,n} (-1)^m \hat a^{\dag}_{m,n} \sigma_z \hat a_{m,n}\notag ,
\end{align}
 where $m$ and $n$ label the horizontal and vertical integer coordinates. The spin-dependent staggered potential could be realized, for instance, by considering an appropriate anti-magic wavelength, for which the polarizability is opposite for the two spin species \cite{gerbier10}. In order to keep the bare tunneling processes of strength $J_{x,y}$ spin-independent the remaining lattice potentials need to be created using a magic wavelength, for which the polarizability is the same for the two spin-species.

 The bare tunneling is suppressed along the legs due to the offset $\Delta \gg J_{x,y}$, which allows for a complete control over induced-tunneling-matrix elements, such as those realized by modulating the lattice resonantly~\cite{goldman14b}.  Following Ref.~\cite{bloch14b}, the modulation is taken to be produced by two pairs of laser beams with frequency difference $\omega_{r/b}=\pm \Delta$ (Fig.~\ref{fig:realization}) in order to restore resonant tunneling. The corresponding time-dependent potential defined by these four lasers is then of the form
\begin{equation}
\hat V(t)= \kappa \sum_{m,n} \hat a^\dag_{m,n}\hat a_{m,n} \left[v(m,n)e^{i \omega t} + v^*(m,n)e^{-i \omega t} \right],\label{mod_pot}
\end{equation}
 with the resonance condition $\omega=\Delta$, and we choose the laser phases in such a way that
\begin{equation}
 v(m,n)=\frac{1}{2}\left\lbrace \cos\left(m\frac{\pi}{2}-\pi/4\right)e^{-i \pi n - i B/2} +  \cos\left(m\frac{\pi}{2}+\pi/4\right)e^{i \pi n + i B/2} \right\rbrace .
\end{equation}
This requires a stabilization of the phase of the modulation relative to the static lattice potential, which is challenging and was not yet demonstrated in previous realizations~\cite{bloch13, ketterle13, bloch14, bloch14b}. This specific choice of the potential $v(m,n)$ is made in order to independently address successive hopping terms along the $x$ direction, which is generally required when engineering Peierls phase-factors in superlattice structures, see Refs.~\cite{goldman14b,bloch14b} and below. 

 The time-evolution of the system is ruled by the Schr\"odinger equation $i \partial_t \psi = \hat H(t) \psi$, where $\hat H (t)=\hat H_0+\hat V(t)$ is defined by Eqs.~\eqref{ham_lab} and~\eqref{mod_pot}. The long-time dynamics of the system can be suitably described by an effective-Hamiltonian approach~\cite{goldman14b}, which is valid in the high-frequency regime $\omega \rightarrow \infty$. Since the static Hamiltonian $H_0$ contains a staggered-potential term that explicitly diverges linearly with $\Delta=\omega$, we first apply the unitary transformation~\cite{goldman14b}
\begin{equation}
\psi =  \hat R(t) \tilde \psi = \exp \left (\! -i  \hat W t \right ) \tilde \psi,\label{transf_hof}
\qquad
\hat W = \frac{\Delta}{2} \sum_{m,n} (-1)^{m} \hat a^\dag_{m,n} \sigma_z \hat a_{m,n}    \, ,
\end{equation}
which removes the diverging term. The effective Hamiltonian can then be derived  in this moving frame, using the method of Refs.~\cite{goldman14b, dalibard14} (see also Ref.~\cite{hauke12}).

 For the sake of simplicity, let us first consider  the dynamics associated with the species $\sigma_z=+1$. For these atoms, the transformed Hamiltonian reads:
\begin{equation} \label{ham_dec}
 \tilde H(t) = \hat R^{\dagger}(t) \left [\hat H_0 + \hat V(t) \right ] \hat R(t) - \hat W= \hat V^{+} e^{i\omega t} + \hat V^{-}  e^{-i\omega t},
\end{equation}
where 
\begin{align}
& \hat V^{+} = \kappa \sum_{m,n} \hat n_{m,n} v(m,n) -J_x \sum_{m \, {\rm odd},n} \left(\hat a^\dag_{m+1,n}\hat a_{m,n} + \hat a^\dag_{m-1,n}\hat a_{m,n} \right) \,, \notag \\
& \hat V^{-} = \kappa \sum_{m,n} \hat n_{m,n} v^*(m,n) - J_x \sum_{m \, {\rm even},n} \left(\hat a^\dag_{m+1,n}\hat a_{m,n} + \hat a^\dag_{m-1,n}\hat a_{m,n} \right) \,.\label{theVeq}
\end{align}
We describe the time-evolution of the system dictated by $\tilde H(t)$ through the evolution operator, which we partition as 
\begin{align}
\tilde U(t) =e^{-i \hat K(t)}e^{- i t \hat H_{\rm eff}}e^{i \hat K(0)},\label{partition} 
\end{align}
where the effective Hamiltonian $\hat H_{\rm eff}$ describes the long-time dynamics, and where the operator $\hat K(t)$ captures the micro-motion. Following Ref.~\cite{dalibard14}, we find that the effective Hamiltonian associated with the general time-dependent Hamiltonian in Eq.~\eqref{ham_dec} is given by: 
\begin{align}
\hat H_{\rm eff}&=\frac{1}{\omega }  [\hat V^{(+1)} , \hat V^{(-1)} ] + \mathcal{O} (1/\omega^2) \label{explicit_eff} \\
 &= -\frac{J_x \kappa}{\omega} \left[ \sum_{m\, \rm{even},\,n} \hat a^\dag_{m+1,n}\hat a_{m,n} \left(v(m+1,n) - v(m,n) \right)\right.+ \nonumber\\
 & \qquad \qquad \quad \left. \phantom{\sum_{n}} +  \hat a^\dag_{m-1,n}\hat a_{m,n} \left(v(m-1,n) - v(m,n) \right)  \right] + \text{H.c.} \nonumber\\
&= \frac{J_x \kappa}{\sqrt{2}\omega} \left[ \sum_{m\,\rm{even},\,n} \hat a^\dag_{m+1,n} \hat a_{m,n} \cos \frac{m\pi}{2} e^{i n\pi + i B/2}   + \hat a^\dag_{m-1,n}\hat a_{m,n} \cos \frac{m\pi}{2} e^{-i n\pi - i B/2}  \right] + \text{H.c.} \nonumber\\
&=\frac{J_x \kappa}{\sqrt{2}\omega} \sum_n  e^{i n\pi + i B/2} \Biggl[ \sum_{m\,\rm{even}}  \left( \hat a^\dag_{2m+1,n}\hat a_{2m,n}+ \hat a^\dag_{2m,n}\hat a_{2m-1,n}\right) \notag \\
& \qquad \qquad \qquad \qquad + \sum_{m\,\rm{odd}}  \left( -\hat a^\dag_{2m+1,n}\hat a_{2m,n} - \hat a^\dag_{2m,n}\hat a_{2m-1,n}\right)  \Biggr]+ \text{H.c.} \notag
\end{align}
The irrelevant sign change in the tunneling matrix elements (i.e. in the last line of Eq.~\eqref{explicit_eff}), can be removed by applying an additional gauge transformation 
\begin{equation}
 \hat G=\exp\left[ i\pi \sum_{m \, \text{odd},n} \hat a^{\dag}_{2m,n}\hat a_{2m,n}\right]= \hat G^{\dagger}, \quad \hat G^2=1\,,
\end{equation}
which indeed reverses the sign of the tunneling terms $\hat a^\dag_{2m+1,n}\hat a_{2m,n}$ and $\hat a^\dag_{2m,n}\hat a_{2m-1,n}$ for $m$ odd only. In this way, the final effective Hamiltonian describing the dynamics of the $\sigma_z=+1$ species  yields
\begin{equation}
 \hat H_{\rm eff}\to \hat G \hat H_{\rm eff}\hat G =\frac{J_x \kappa}{\sqrt{2}\omega} \sum_{m,n}  \hat a^\dag_{m+1,n}\hat a_{m,n} e^{i n\pi + iB/2}  + \text{H.c.}
\end{equation}
which is indeed the tunneling term in Eq.~\eqref{ham} for $\sigma_z=+1$ atoms.
 In the case of the $\sigma_z=-1$ species, the staggered potential is reversed, so that the even and odd sites must be inverted. This results in the final effective Hamiltonian:
\begin{equation}
 \hat H_{\rm eff}=\frac{J_x \kappa}{\omega} \sum_{m,n} \left( (-1)^n\hat a^\dag_{m+1,n} e^{i B\sigma_z/2} \hat a_{m,n}   + \text{H.c.} \right)\,,
\end{equation}
where $(-1)^n$  is equivalent to the operator $\tau_z$ in Eq.~\eqref{ham}.

The Zeeman term $\Omega \sigma_x$ present in the Hamiltonian~\eqref{Hamspace} can be directly generated by two resonant coupling potentials $\hat{V}^{RC}_i(t)=\sum_{m,n} 2\Omega \cos (\nu_i t) \hat{a}_{m,n} \sigma_x a_{m,n}$.  Indeed, considering the bare atomic frequency $\omega_z \gg \Delta$ between the two sublevels, they are effectively separated by the position dependent energy offset $\omega_z +(-1)^m\Delta$ created by the spin-dependent potential in Eq.~\eqref{ham_lab}. Under the transformation $\hat{R}(t)$, the effect of the coupling becomes 
\begin{equation} \label{rotzeeman}
 \hat{R}^\dag(t) \hat{V}^{RC}_i(t) \hat{R}(t)  = \sum_{m,n} 2\Omega e^{-i\Delta (-1)^m  t} \cos \left( \nu_i t\right)  \hat{a}^\dag_{m,n} \sigma_+ \hat{a}_{m,n} + \text{H.c.}  
\end{equation}
where $\sigma_+=(\sigma_x+i\sigma_y)/2$. This terms commute with the gauge-transformation operator $\hat{G}$ and, by choosing the frequencies $\nu_{1,2}=\omega_z \pm \Delta$, we recover the required Zeeman term $\Omega \sigma_x$ in Eq.~\eqref{Hamspace},  both for even and odd sites, through the standard rotating-wave approximation.

We note that the effective Hamiltonian~\eqref{Hamspace} was derived at first order in $\omega^{-1},$ in a basis provided by two commuting unitary operators: $\hat{R}(t)$ and $\hat{G}$. It is important to notice that these latter operators commute with the $\tau$ operators, so that they neither affect the static hopping term $J \tau_x$, nor the static potential difference $\mu \tau_z$: this indicates that these static terms can be directly included in the (effective) Hamiltonian~\eqref{Hamspace}. The latter remark is also valid for the spin-independent potential $\mu_0$. Therefore, we conclude that the application of two pairs of Raman lasers and a radio-frequency field, combined with the spin-dependent staggered potential directed along the ladder, allows one to generate all the spin-dependent terms in the ladder Hamiltonian~\eqref{Hamspace}.

Importantly, we emphasize that the Hubbard interactions are also unaffected by the aforementioned transformations $\hat{G}$ and $\hat{R}(t)$. Thus, the effects of interactions can be directly incorporated into the  effective Hamiltonian~\eqref{Hamspace}, at first order in $\omega^{-1}$.

Finally, the time-evolution operator in Eq.~\eqref{partition} is then completely determined by computing the kick operator $\hat{K}(t)$, which is readily calculated using the expression \cite{dalibard14,goldman14b}
\begin{align}
\hat K (t) &= \frac{1}{i  \omega}  \left [\hat V^{+} e^{i \omega t} - \hat V^{-} e^{-i \omega t}     \right ] 
\approx  \frac{2 \kappa}{\omega} \sum_{m,n} \hat n_{m,n}\, \vert v (m,n) \vert  \sin (\omega t + \theta_{m,n}),   
\end{align}
where we assumed that $\kappa \gg J_x$ and we defined $\theta_{m,n} = \text{arg} [v (m,n)]$. 

\section{Conclusions} \label{sec:conclusions}

In this work we presented a ladder setup for ultracold fermions subject to both the presence of an artificial $\pi$-flux magnetic potential and a spin-orbit coupling. Such a model may be seen as the synthesis of two accessible experimental techniques  to realize synthetic gauge fields in optical lattices: on one side, the realization of complex tunneling matrix elements using time-modulated optical lattice \cite{esslinger14,bloch14b,bloch14,bloch13,ketterle13}, and, on the other, the implementation of spin-orbit terms through the control over internal atomic degrees of freedom \cite{mancini_2015,stuhl_2015}.
Analogously to models already discussed in the context of nanowires \cite{loss13,klinovaja13}, the combination of these two elements gives rise to a particle-hole symmetry which  protects non-trivial topological phases  within the Bogoliubov-de Gennes and chiral classes of topological insulators and superconductors.
 As the model conserves the number of particles and can be realized with state-of-the-art experimental techniques, our results might give a substantial advance towards the observation of topologically-protected zero-energy modes in fermionic systems with and without interactions. In particular the physical realization that we present does not require the engineering of any pairing mechanism, neither the coupling to external molecular gases or superfluids, as, for example in \cite{cirac11,kraus12,nascimbene13,buhler14}, nor an interchain pair-hopping, as exploited in the ladder model presented in \cite{kraus13}.
 
No interaction is indeed necessary for the appearance of the symmetry-protected topological phase. Consequently, the realization of the particle-hole symmetry through the ladder geometry is not as robust as its counterpart in topological superconductors due to the absence of a true superconducting gap, thus it should be considered an extrinsic feature. However, due to the absence of disorder and to the high degree of isolation in ultracold atom setups, we expect the topological features of the system to be experimentally detectable. For example, the introduction of a trapping potential does not spoil the observation of edge modes, despite breaking the particle-hole symmetry.

We analyzed two experimentally relevant signatures of the appearance of topological phases: the presence of fractionalized edge modes, detectable through site-resolved density measurements (as recently reported in Refs.~\cite{kuhr2015,zwierlein15, greiner2015}), and the winding behavior of the spin degree of freedom, which can be observed through spin-resolved time-of-flight imaging.
In particular, we have shown how to detect these observables even in the presence of a trapping potential, which induces soft boundaries, often believed to be particularly disruptive for the detection of topological signatures.

Our study has also considered the effect of a contact repulsive interaction; this is possible thanks to the presence of the spin degree of freedom that differentiate the main features of our model from its spinless counterparts as the SSH and its interacting generalizations (see for example \cite{li2013,grusdt2013}). Apart from mapping out the phase-diagram of the model, which entails two gapped phases, with and without topological properties, we have found that a Hubbard interaction can enhance the extension of the topological phase, instead of being detrimental. Furthermore, the spin winding number introduced in the article provides a good topological parameter also in the interacting case, where the usual order parameters based on single-particle wavefunctions fail.

Concluding, we mention that the model under scrutiny may be an interesting platform for the study of further fractionalization effects, based on a particular fine tuning of the parameters, which can be reminiscent of the physics of parafermionic zero-energy modes (see Ref.~\cite{alicea15} for a recent review) in the spirit of Ref.~\cite{klinovaja13}.

\section*{Acknowledgements}

We acknowledge enlightening discussions with Rosario Fazio, Fabian Hassler, Sylvain Nascimbene, Luca Taddia and Andrea Trombettoni and thank Simone Barbarino for careful reading of the manuscript. We thank Davide Rossini for providing the numerical algorithms based on Matrix-Product States employed in this work. 
L.~M.~acknowledges support from the Italian MIUR through FIRB project RBFR12NLNA and Regione Toscana POR FSE 2007-2013. 
M.~A., H.-H.~T. and M.~B.~acknowledge support from the EU grant SIQS.
N.G. is financed by the FRS-FNRS Belgium and by the BSPO under the PAI project P7/18 DYGEST. 

\appendix

\section{Symmetries of the model, order parameter and spin winding number} \label{app:spin}

In this Appendix we examine the non-interacting ladder model and we discuss, in particular, the relation between the order parameter $\mathcal{W}$, that distinguishes trivial and topological phases, and the observed spin winding number.

The Hamiltonian \eqref{Hamspace} is translationally invariant and can be also expressed in momentum space as $\hat H = \sum_k \hat a^\dagger_k \mathcal H(k) \hat a_k$ where:
\begin{equation} \label{ham}
\mathcal H(k)= 2t\tau_z\cos(B/2)\cos k - 2t \tau_z\sigma_z \sin(B/2)\sin k + \Omega\sigma_x + J\tau_x +\mu \tau_z + \mu_0.
\end{equation}
We observe that its kinetic term corresponds to a Peierls substitution $k \to k + \frac{B\sigma_z}{2}$.

Its spectrum is symmetric for a tranformation mapping $B \to B+2\pi$ and $t \to -t$. Therefore, even if the Hamiltonian is periodic in $B$ with period $4\pi$, we can restrict our study to the case $0\le B <2\pi$. Besides, we can consider only positive values of $\Omega$ and $J$ because their sign trivially depends on the chosen basis for the spin and pseudospin. In particular $\mathcal H(k,-\Omega)=\sigma_z \mathcal H(k,\Omega) \sigma_z$ and $\mathcal H(k,-J)=\tau_z \mathcal H(k,J) \tau_z$. Analogously we consider only $t>0$, because of the symmetry between the two chains $\mathcal H(k,-t,-\mu)=\tau_x\mathcal H(k,t,\mu)\tau_x$. Hereafter we rescale all the energies in units of $t$ in such a way that, below, we will always consider $t=1$. Finally we observe that the expectation value of $\sigma_y$ is always null because the Hamiltonian \eqref{ham} is real.

The system is characterized by an anti-unitary time-reversal-like symmetry $T=\sigma_x$,
\begin{equation}
 T\mathcal H(k)T^\dag=\mathcal H^*(-k)\,,
\end{equation}
and, for $\mu_0=0$ (the system is exactly at half filling), we obtain the particle-hole symmetry $C=\tau_y\sigma_y$:
\begin{equation}
 C\mathcal H(k)C^\dag=-\mathcal H^*(-k).
\end{equation}
These non-unitary symmetries characterize the topological symmetry class BDI (see, for example, \cite{kitaev09,ludwig08}), which is also characterized by the unitary chiral symmetry $P=TC=\tau_y\sigma_z$.

Additional terms in the Hamiltonian may break the $C$ symmetry which is indeed fragile, due to the lack of a physical pairing interaction; we emphasize however that such perturbations become significant only if their magnitude is comparable with the energy gap. To this concern, in a cold atom gas, the presence of noise and defects are negligible and the main effect we must consider is the trapping potential bringing to a space dependent chemical potential $\mu_0$. As discussed in Sec. \ref{sec:Topological}, however, the local shift in energy provided by the trapping allows in general to isolate topological regions of the chain with the effect of binding fractionalized modes at the interface between these regions and the trivial ones.

An additional Zeeman term proportional to $\sigma_z$, which breaks only the time reversal symmetry, brings instead the system in the symmetry class $D$, which is still topologically non-trivial in one dimensions. Therefore, this sort of term does not alter in a fundamental way the properties of the system.

One-dimensional systems in the BDI class possess topological phases labelled by a topological invariant in $\mathbb{Z}$ \cite{ludwig08}. This topological invariant, can be evaluated by exploiting the chiral basis defined by the symmetry $P$ \cite{sau12}. In this basis the Hamiltonian assumes the simple form:
\begin{equation} \label{hamchiral}
 \mathcal H(k)=\begin{pmatrix}
      0 & A(k) \\ A^{\dag}(k) & 0
     \end{pmatrix};
     \;
A=
     \begin{pmatrix}
 i \mu +J +2 i  \cos \left(\frac{B}{2}+k\right) & i \Omega  \\
 i \Omega  & i \mu -J +2 i  \cos \left(\frac{B}{2}-k\right) \\
\end{pmatrix}.
\end{equation}

If $\det(A(k))\ne 0$ for all $k$ then the system is gapped. In this case, $\det{A(k)} = \left| \det{A(k)}\right| e^{i\xi(k)} $ and the winding number of $\xi(k)$ constitutes the topological invariant which distinguish topological and non-topological phases \cite{sau12}. In this model, this winding number may assume only the values $0$ or $\pm 1$ characterizing trivial and topological phases respectively. In more detail, one obtains:
\begin{multline}
 \det A(k)= -J^2 -\mu^2 + \Omega^2 -2\cos(B) -2\cos(2k)+ \\ -4\mu\cos(B/2)\cos(k)  + 4iJ\sin(B/2)\sin(k)\,.
\end{multline}
This determinant is purely real for $k=0,\pi$ or for $B=0$. For $B=0$, its phase $\xi(k)$ cannot change its value from $0$ to $\pi$ unless $\det A$ crosses zero and the gap closes; thus the case $B=0$ is either trivial or gapless.

\begin{figure}[t]
 \centering
 \includegraphics[width=0.6\columnwidth]{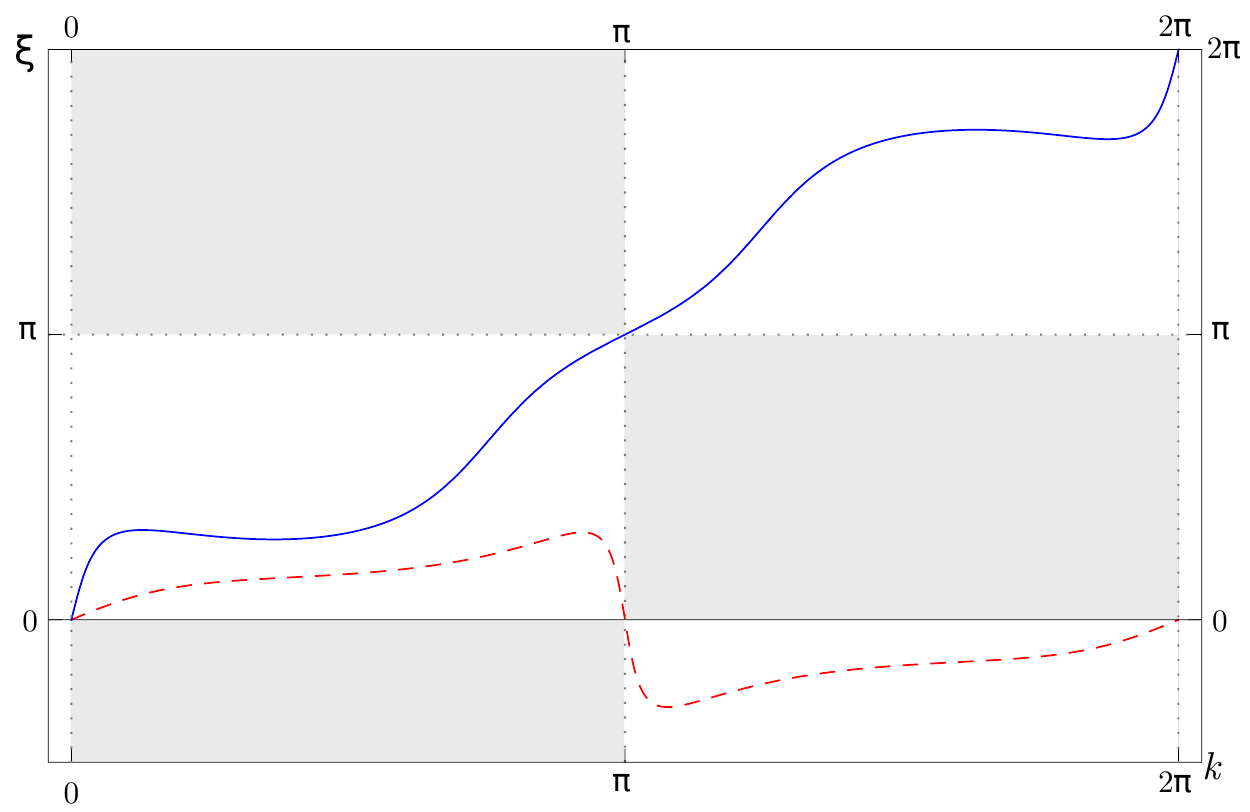}
 \caption{Different behavior of the phase $\xi$ as a function of the momentum $k$ for a trivial phase (red dashed line) and the topological phase (blue line). In the trivial case the values of $\xi$ at $k=0,\pi,2\pi$ are equal and the winding number parity is $\mathcal{W}=1$; for the topological phase, instead, $\xi(k=0)$ and $\xi(k=2\pi)$ differ by $2\pi$. The phases were calculated for $\mu=J=1$ and $B=4\pi/3$ for $\Omega=1.2$ in the topological region and $\Omega=2.3$ in one of the trivial regions. The shaded regions are forbidden for $B\in(0,2\pi)$. } \label{wind}
\end{figure}

In all the other cases, for $0<B<2\pi$, the topological invariant $\mathcal{W}$ is evaluated by considering the behavior of $\xi$ between $k=0$ and $k=\pi$. In particular, the tangent of the phase $\xi$ is given by:
\begin{equation} \label{tangent}
 \tan\left[\xi(k) \right] =
 \frac{4 J \sin \frac{B}{2} \sin k}{\Omega ^2-4 \mu  \cos \frac{B}{2} \cos k-2 \cos B-J^2-2 \cos (2 k)-\mu ^2}
\end{equation}
where the numerator is always positive for $B\in(0,2\pi)$ and $k \in (0,\pi)$, and always negative for $B\in(0,2\pi)$ and $k \in (\pi,2\pi)$.
Therefore, going from $k=0$ to $k=\pi$, $\xi(k)$ must be always included in $(0,\pi)$ since $\sin(\xi)>0$ in this regime, whereas for $k$ which goes from $\pi$ to $2\pi$, $\xi(k)$ must be either in $(-\pi,0)$ (if $\xi(k=\pi)=0$), such that its winding number vanishes for $k\to 2\pi$, or $\xi(k)\in(\pi,2\pi)$ which implies a final winding number equal to one (see Fig. \ref{wind}). Thus the parity of the winding number results:
\begin{equation} \label{winv}
 \mathcal{W}= {\rm sign}\left[  \frac{\Omega^2 -2-J^2-\mu^2-4\mu\cos(B/2)-2\cos(B)}{\Omega^2-2-J^2-\mu^2+4\mu\cos(B/2)-2\cos(B)}\right].
\end{equation}

$\mathcal{W}=1$ when the winding number is $0$, whereas $\mathcal{W}=-1$ for the winding number being $\pm 1$. Therefore, in terms of the parameter $\Omega$, two phase transitions appear at:
\begin{equation} \label{trans1}
\Omega^2_{c,1} \equiv J^2+\min\left[\left( \mu - 2\cos\frac{B}{2}\right)^2, \left( \mu + 2\cos\frac{B}{2}\right)^2 \right]
\end{equation}
and
\begin{equation} \label{trans2}
\Omega^2_{c,2} \equiv J^2+\max\left[\left( \mu - 2\cos\frac{B}{2}\right)^2, \left( \mu + 2\cos\frac{B}{2}\right)^2 \right].
\end{equation}
The system is in a topological phase $(\mathcal{W}=-1)$ for $\Omega^2_{c,1}< \Omega^2 < \Omega^2_{c,2}$ whereas for $\Omega^2<\Omega^2_{c,1}$ and $\Omega^2>\Omega^2_{c,1}$ we obtain trivial phases $(\mathcal{W}=1)$. At the transition points $\Omega^2=\Omega^2_{c,1},\Omega^2_{c,2}$, the gap closes respectively for $k=\pi,0$, consistently with the results in \cite{klinovaja13,oreg14}. The corresponding critical values for $J$ are given by:
\begin{equation}
 J^2_{1,2} = \Omega^2 - \left( \mu \mp 2\cos(B/2)\right)^2
\end{equation}
Depending on $\Omega, \mu$ and $B$ there can be 0,1 or 2 phase transitions as a function of $J$.

In order to evaluate the spin expectation value in the plane $\hat{x}-\hat{z}$, it is useful to adopt the Hamiltonian form in Eq. \eqref{hamchiral} which allows a simple diagonalization. In this basis the components of the physical spin, $S^x$ and $S^z$, become:
\begin{equation}
 S^x = \sigma_x \otimes \tau_y \,,\qquad S^z=\sigma_z \otimes \tau_0.
\end{equation}
where $\sigma_i$ and $\tau_i$ are Pauli matrices in this basis.
To obtain the eigenstates of the Hamiltonian we consider:
\begin{equation}
 \mathcal H^2(k)=\begin{pmatrix}
      A(k)A^{\dag}(k) & 0 \\ 0 &  A^{\dag}(k)A(k)
     \end{pmatrix}
\end{equation}
where, $AA^\dag=(A^\dag A)^*$ since $A=A^T$. Therefore the generic form of the eigenstates in this basis is:
\begin{equation}
\Psi = \begin{pmatrix} \psi \\ \pm e^{i\alpha} \psi^* \end{pmatrix}
\end{equation}
where $\psi$ are the two eigenvectors of $AA^\dag$. In particular, we are interested in the eigenvector corresponding to the lowest eigenvalue of  $AA^\dag$, since it determines the second and the third bands of $H$ which are topologically non-trivial. The value of $\alpha$ is fixed by the equation:
\begin{equation}
\psi^\dag A \psi^* = \pm \varepsilon e^{-i\alpha}
\end{equation}
where $\pm \varepsilon$ are the energies of the two intermediate bands of the system.

The expectation value of $S^z$ in the second band does not depend on the phase $\alpha$ and it is easily evaluated. It results:
\begin{equation}
 S^z_2(k)\equiv \left\langle S^z \right\rangle =\frac{2 \sin (k) \left(\sin (B) \cos (k)+ \mu \sin \left(\frac{B}{2}\right)\right)}{N(k)}
\end{equation}
where $N(k)$ is a positive quantity. In particular we must distinguish two cases: $|\mu| \gtrless \left|\cos{B/2}\right|$.

For $\mu > 2\left|\cos{B/2}\right|$, $S^z_2>0$ for $k\in(0,\pi)$, thus $S^z_2$ behaves like the imaginary component of the phase $\exp[i\xi(k)]$ and it can be showed that the sign of $S^x_2$ at $k=0$ and $k=\pi$ is equal to the sign of the denominator of Eq. \eqref{tangent}, therefore the spin winding number of the second band and the winding number of the phase $\xi(k)$ coincide.

For $0 < \mu< 2\left|\cos{B/2}\right|$, $S^z_2(k)$ has a further zero between $k=0$ and $k=\pi$, therefore the spin vector is aligned along $\hat{x}$ in three different points in the interval $k\in[0,\pi]$. This can be seen in the last column of Fig \ref{fig:trap} (obtained for $\mu=0.5$ and $B=\pi/2$), where the blue line, representing the average value of the spin in a system without the trapping, always crosses the horizontal axis three times. The spin winding number $\mathcal S_2$ is null if and only if $S^x_2(k)$ has the same sign in these three points (first and last rows). However if $S^x_2(k)$ assumes the same sign for $k=0$ and $k=\pi$, then the total spin winding number for $k$ going from $0$ to $2\pi$ is even, due to the time-reversal symmetry $T$. Otherwise it is odd. At $k=0$ and $k=\pi$ the sign of $S^x_2(k)$ assumes, respectively, the same value of the signs of the numerator and denominator of $\mathcal{W}$ in Eq. \eqref{winv}. Therefore the parity of the spin winding number, $(-1)^{\mathcal{S}_2}$, always coincides with $\mathcal{W}$.

Concerning the behavior at the intermediate zero of $S^z_2$,
\begin{equation}
 k_c=\arccos\left(-\mu \sin(B/2)/\sin(B) \right),
\end{equation}
one has $S^x_2(k_c) = \cos\left[ \alpha(k_c)\right] \propto \Omega-J$, therefore in the trivial region where $\Omega^2>\Omega_{c,2}^2$ the spin winding number is trivial. The other trivial region, $\Omega^2<\Omega_{c,1}^2$, is instead divided into two subregimes characterized by a spin winding number $\mathcal S_2=0$ or $\mathcal S_2 =2$ separated in $J=\Omega$. This is shown in the last column of Fig. \ref{fig:trap} where the blue lines either does not wind around the origin (first row, corresponding to $\Omega < J<\Omega_{c,1}$) or winds twice around it (second row, $\Omega_{c,1}>\Omega>J$).

Similar conclusions hold for negative values of $\mu$ and it can be shown that the behavior of the spin winding number is not affected by open or periodic boundary conditions.

\end{document}